# The sensitivity of a radical pair compass magnetoreceptor can be significantly amplified by radical scavengers

Daniel R. Kattnig[1,2] & P. J. Hore[1]

**Birds have a remarkable ability to obtain navigational information from the Earth's magnetic field. The primary detection mechanism of this compass sense is uncertain but appears to involve the quantum spin dynamics of radical pairs formed transiently in cryptochrome proteins. We propose here a new version of the current model in which spin-selective recombination of the radical pair is not essential. One of the two radicals is imagined to react with a paramagnetic scavenger via spin-selective electron transfer. By means of simulations of the spin dynamics of cryptochrome-inspired radical pairs, we show that the new scheme offers two clear and important benefits. The sensitivity to a 50 μT magnetic field is greatly enhanced and, unlike the current model, the radicals can be more than 2 nm apart in the magnetoreceptor protein. The latter means that animal cryptochromes that have a tetrad (rather than a triad) of tryptophan electron donors can still be expected to be viable as magnetic compass sensors. Lifting the restriction on the rate of the spin-selective recombination reaction also means that the detrimental effects of inter-radical exchange and dipolar interactions can be minimised by placing the radicals much further apart than in the current model.**

Magnetoreception — the ability to sense magnetic fields — is widespread throughout the animal kingdom[1], but the underlying detection mechanisms are far from clear[2]. There are two main hypotheses. One involves single-domain, ferrimagnetic or superparamagnetic iron-containing particles that are caused to move by their interaction with the Earth's magnetic field and so influence the gating of mechano-sensitive or force-gated ion channels[3–7]. The other is based on radical pairs formed by photo-induced electron transfer reactions in sensor proteins[2,8,9]. The magnetic sensitivity arises from a combination of the coherent quantum spin dynamics and the spin-selective reactivity of a pair of spin-correlated radicals, which cause the yield of a signalling state to depend on the intensity and direction of the external magnetic field.

The radical pair mechanism is "quantum" not just in the trivial sense that it involves a non-classical property (spin angular momentum), but more interestingly because quantum coherences play an essential role[10]. This aspect was highlighted by a recent suggestion that the angular precision of the magnetic compass in migratory birds can be understood in terms of avoided crossings of spin energy-levels in the radicals[11]. Another quantum feature of the mechanism is that the electron spins in the radical pair are initially entangled although there is currently no reason to think that entanglement, as distinct from coherence, is necessary for the operation of the compass[12–14]. It is also striking that the performance of a radical pair sensor may be enhanced by interactions with its fluctuating environment[15], a property apparently found in other areas of "quantum biology"[16–19].

Experimental and theoretical support for a radical pair mechanism of magnetoreception is accumulating (reviewed in ref. 2), in particular in the context of the avian magnetic compass. Magnetically sensitive radical pairs are thought to be formed by light-dependent electron transfer reactions in cryptochromes — blue-light photoreceptor flavoproteins — located in the retina[20,21]. *In vitro*, purified cryptochrome from the plant *Arabidopsis thaliana* (*At*Cry1) and a closely related protein, *E. coli* photolyase (*Ec*PL), both show light-dependent responses to weak magnetic fields (~1 mT)[22,23]. In these proteins, photo-excitation of the non-covalently bound flavin adenine dinucleotide (FAD) cofactor leads to the formation of radical pairs via sequential electron transfers along the "tryptophan-triad", a chain of three conserved tryptophan residues within the protein[24–26] (Fig. 1). This process reduces the photo-excited singlet state of the FAD to the anion radical, $FAD^{\bullet-}$, and oxidises the terminal, surface-exposed, tryptophan ($Trp_CH$) to give the cation radical, $Trp_CH^{\bullet+}$. Formed with conservation of spin angular momentum, the radical pair is initially in an electronic singlet state, $^1$[$FAD^{\bullet-}$ $Trp_CH^{\bullet+}$][22,23,27]. This form of the protein is a coherent superposition of the eigenstates of the spin Hamiltonian which comprises the Zeeman, hyperfine, exchange and dipolar interactions of the electron spins. As a consequence, the radical pairs oscillate coherently between the singlet and triplet states, a process that manifests itself in the yields of subsequent spin-selective reactions of the radicals. In particular, when the protein is immobilized, the anisotropy of the electron-nuclear hyperfine interactions causes the reaction product yields to depend on the orientation of the protein with respect to an external magnetic field.

[1]Department of Chemistry, University of Oxford, Physical and Theoretical Chemistry Laboratory, South Parks Road, Oxford, OX1 3QZ, UK. [2]Present address: Living Systems Institute and Department of Physics, University of Exeter, Stocker Road, Exeter, EX4 4QD, UK. Correspondence and requests for materials should be addressed to P.J.H. (email: peter.hore@chem.ox.ac.uk)

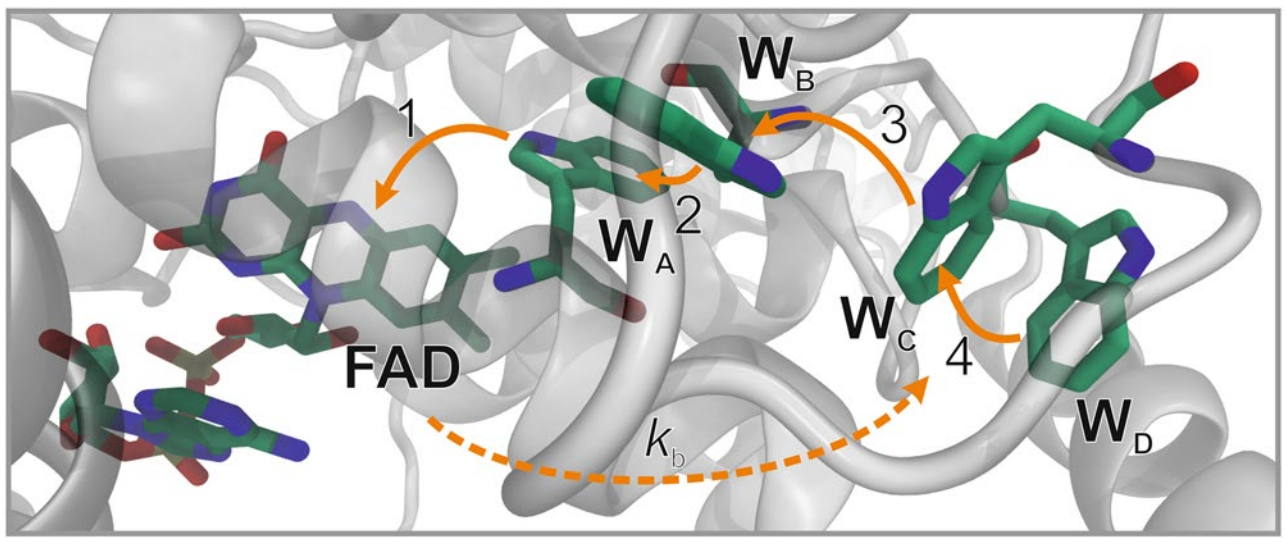

**Figure 1.** Electron transfer pathway in cryptochromes. After photo-excitation of the FAD cofactor, three or four rapid sequential electron transfers along a triad or tetrad of tryptophan residues ($W_A$, $W_B$, $W_C$, $W_D$) generate a spin-correlated radical pair [$FAD^{\bullet-}$ $Trp_CH^{\bullet+}$] or [$FAD^{\bullet-}$ $Trp_DH^{\bullet+}$]. The figure is based on the crystal structure of *Dm*Cry (PDB ID: 4GU5)[35,36].

Following Ritz *et al.*[9], most authors (e.g. refs 14, 28–33) have envisaged spin-selective reactions of both singlet and triplet radical pairs. The latter requires there to be a triplet product state that is energetically accessible from the radical pair. As no such species exists in cryptochrome, we base our treatment here on the more plausible scheme shown in Fig. 2a which we henceforth refer to as the "current" model[22,23,34]. In Fig. 2a, the spin-selective reaction channel is charge recombination (rate constant $k_b$) within the singlet state of the radical pair (**RP**) which regenerates the ground state of the protein, **G**. As found experimentally for the flavin-tryptophan radical pair in *At*Cry1 and *Ec*PL[22,23], the **RP** state also undergoes proton transfer reactions, which occur with equal rate constants ($k_f$) for singlet and triplet pairs, to produce a secondary, long-lived radical pair state, **S**. *In vivo*, **S** is assumed to be, or to lead to, the biochemical signalling state of the protein[2]. The interaction of the electron spins with the magnetic field can induce a significant change in the yield of **S** if $k_b^{-1}$ (the characteristic time of singlet recombination) is comparable to or shorter than $k_f^{-1}$ (the time required for the formation of **S**), small compared to the electron spin relaxation time (~1 μs or possibly longer) and longer than the coherent singlet-triplet interconversion time. These conditions mean that the radicals must not be too far apart, otherwise charge recombination will be too slow. This restriction is satisfied for the *Ec*PL and *At*Cry1, in which the edge-to-edge separation of the aromatic rings of FAD and $Trp_CH$ is ~1.47 nm[35–37]. In the cryptochrome from the fruit fly, *Drosophila melanogaster* (*Dm*Cry), however, there is an additional electron donor, $Trp_DH$, beyond $Trp_CH$ (Fig. 1)[38]. The edge-to-edge distance between FAD and $Trp_DH$ in *Dm*Cry is 1.70 nm[35,36] which is large enough that direct charge recombination in $^1$[$FAD^{\bullet-}$ $Trp_DH^{\bullet+}$] cannot compete effectively with electron spin relaxation, at least for the purified protein *in vitro*, explaining the weak magnetic field effects observed for *Dm*Cry[39]. Sequence alignments suggest that avian cryptochromes also have a fourth tryptophan which could be involved in radical pair formation; we return to this point below.

All of the above has been gleaned from spectroscopic observations of purified cryptochromes *in vitro*[22,23,39,40]. The same proteins may behave differently in a cellular environment as a result of their interactions with (for example) signalling partners and whatever structures immobilise and align them as direction sensors. No studies of magnetic field effects have yet been reported for any of the four avian cryptochromes. There is therefore scope to speculate about alternative radical pairs that might undergo different photo-reactions to those in Fig. 2a[41].

Here we describe a modified reaction scheme in which the magnetic compass sensitivity is enhanced by a spin-selective reaction of one of the constituents of the radical pair with an additional paramagnetic molecule. We refer to this molecule as a 'scavenger', a term defined as "a substance that reacts with (or otherwise removes) a trace component […] or traps a reactive reaction intermediate"[42]. The scavenging process can lead to large magnetic field effects, even in the limit of very slow charge recombination ($k_b \to 0$) where the current model (Fig. 2a) predicts no magnetic field effects at all. Such a reaction scheme could allow radical pairs with separations much larger than 2 nm to operate as efficient magnetic compass sensors including, but not restricted to, those containing a fourth tryptophan residue in the electron transfer chain.

**Spin-selective scavenging.** We start by showing how a spin-conserving reaction with a paramagnetic scavenger can affect the coherent spin dynamics of a radical pair. Our treatment is a generalization of Letuta and Berdinskii[43]. We consider a pair of radicals (A and B) in a spin-correlated singlet state and determine how the spin state changes when one of the radicals (A) undergoes spin-selective reactions with a paramagnetic scavenger C. Molecules A, B and C have spin quantum numbers, $S$, equal to ½, ½, and $J \geq$ ½, respectively. The scavenging reactions convert A into a diamagnetic species ($S=0$), for example by the transfer of an electron to or from C. We focus on organic molecules with weak spin-orbit coupling such that the reactions conform to Wigner's spin conservation rule[44,45]; as a consequence, the spin of C must change by ±½. Thus, in general, there are two parallel, spin-allowed reactions with distinct rate constants, $k_C^{\pm}$, and distinct products:

$$^2\mathrm{A} + {}^{2J+1}\mathrm{C} \xrightarrow{k_C^+} {}^1\mathrm{A} + {}^{2J+2}\mathrm{C} \tag{1}$$

$$^2\mathrm{A} + {}^{2J+1}\mathrm{C} \xrightarrow{k_C^-} {}^1\mathrm{A} + {}^{2J}\mathrm{C} \tag{2}$$

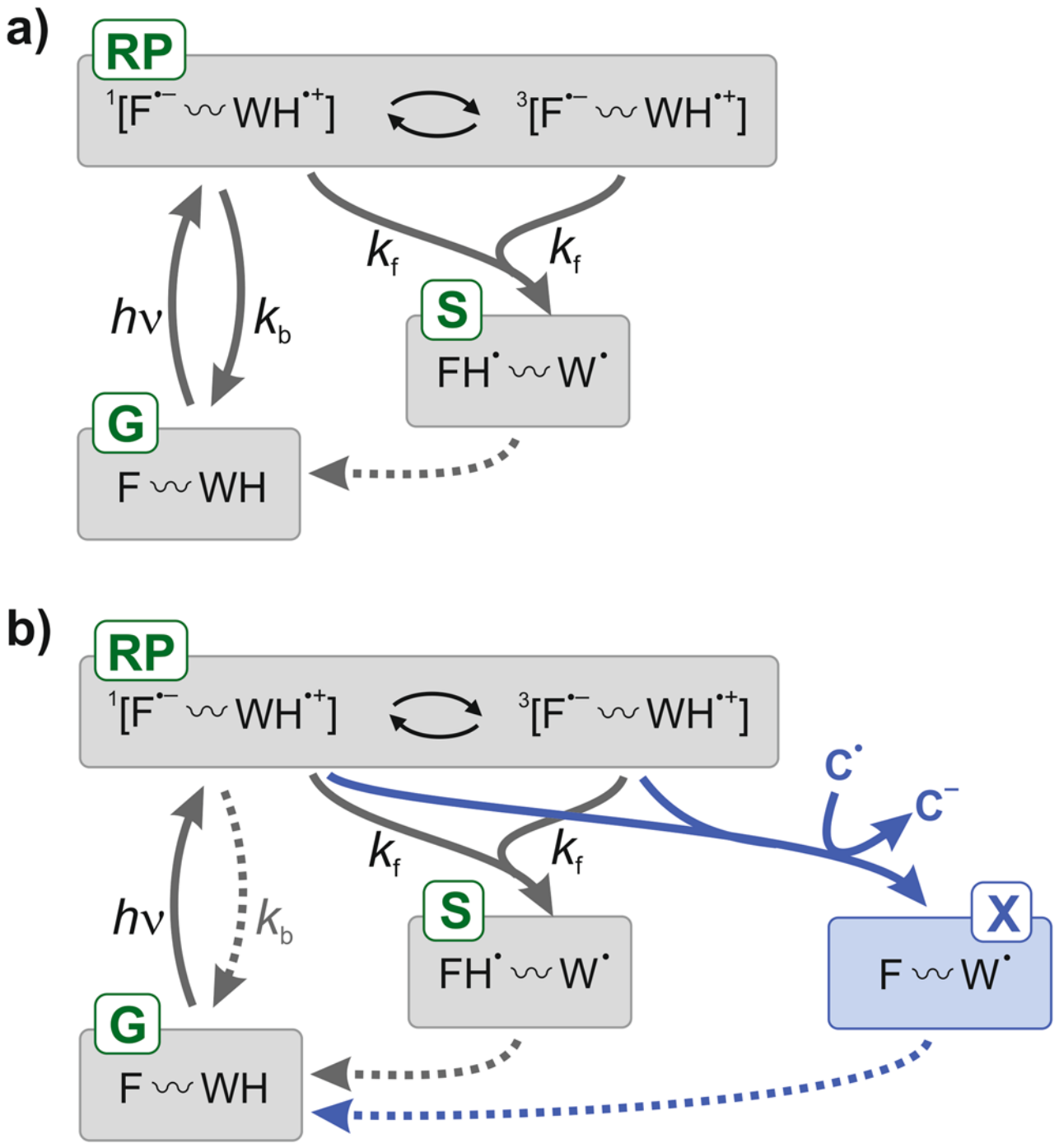

**Figure 2.** Cryptochrome reaction schemes for magnetoreception. (**a**) The photocycle that accounts for the magnetic field effect on *At*Cry1[23]. (**b**) The same reaction scheme augmented by a spin-selective reaction of the flavin radical with a scavenger, C. Abbreviations used for different states of the protein are: **RP**, radical pair state; **G**, ground state; **S**, signalling state; **X**, scavenging product state. Abbreviations used for reaction partners: F, flavin adenine dinucleotide; WH, terminal residue of the Trp triad/tetrad. Superscript dots indicate radicals. Superscript numbers are spin multiplicities. FH$^•$ and W$^•$ are (de)protonated forms of the initially formed radicals, F$^{•-}$ and WH$^{•+}$. The dashed arrows indicate processes that regenerate **G**, typically on a slow timescale, but which are not essential for the function of the sensor. The photo-excited singlet state of the FAD is not shown.

(the superscripts are the spin multiplicities, $2S+1$).

In the combined Hilbert spin-space of A, B and C, these reactions are governed by projection operators that can be written in terms of the spin angular momentum operators, $\hat{\mathbf{S}}_K$ (see Supporting Information, Section S1):

$$\hat{P}_{AC}^{\pm} = \frac{1}{2} \pm \frac{1}{2J+1}\left[\frac{1}{2} + 2\hat{\mathbf{S}}_A \cdot \hat{\mathbf{S}}_C\right], \tag{3}$$

where $\hat{P}_{AC}^{+} + \hat{P}_{AC}^{-} = \hat{1}$ and $(\hat{P}_{AC}^{\pm})^2 = \hat{P}_{AC}^{\pm}$. Following Haberkorn[46], the reactions give rise to the following equation of motion for the (concentration-weighted) density operator of the {A,B,C} spin system:

$$\frac{\mathrm{d}}{\mathrm{d}t}\hat{\rho}(t) = -\frac{1}{2}k_C^+[\hat{P}_{AC}^+, \hat{\rho}(t)]_+ - \tfrac{1}{2}k_C^-[\hat{P}_{AC}^-, \hat{\rho}(t)]_+, \tag{4}$$

where $[\ ,\ ]_+$ denotes the anti-commutator.

To see most clearly the effect these reactions have on the surviving AB radical pairs, we temporarily ignore all spin interactions and all other reaction steps. We assume that the radical pair is initially in a singlet state, specified by the projection operator $\hat{P}_{AB}^{S}$, and that C has no initial spin-correlation with A or B:

$$\hat{\rho}(0) = \hat{P}_{AB}^{S}/\mathrm{Tr}\left[\hat{P}_{AB}^{S}\right] = \left(\tfrac{1}{4} - \hat{\mathbf{S}}_A \cdot \hat{\mathbf{S}}_B\right)/\mathrm{Tr}\left[\hat{P}_{AB}^{S}\right]. \tag{5}$$

Using equations (3)–(5), the fraction of the AB radical pairs that remain unreacted at time $t$ is the sum of two exponential decays (see Supporting Information, Section S2):

$$s(t) = \mathrm{Tr}[\hat{\rho}(t)] = \frac{J+1}{2J+1}\exp(-k_{\mathrm{C}}^{+}t) + \frac{J}{2J+1}\exp(-k_{\mathrm{C}}^{-}t), \tag{6}$$

in which the two terms are the probabilities of the $S = J \pm ½$ coupled angular momentum states of the AC subsystem.

We now make the simplifying assumption that, for energetic reasons, the spin-selective reaction that produces C in its higher spin state ($S = J + ½$) is negligibly slow, i.e. $k_{\mathrm{C}}^{+} = 0$. For example, if C in equation (2) is a doublet (i.e. a radical, with $J = ½$), it can end up as either a singlet ($^1$C, $S = J - ½ = 0$) or a triplet ($^3$C, $S = J + ½ = 1$). For organic molecules, the former is normally the ground state and the latter an excited state. Similarly if C is a triplet ($J = 1$), we consider only the formation of the doublet product, $^2$C, and exclude the higher energy quartet product, $^4$C. With this simplification, therefore, a fraction

$$s(t \to \infty)|_{k_{\mathrm{C}}^{+}=0} = \frac{J+1}{2J+1} \tag{7}$$

of the AB pairs is unreactive. In general, the probability that the surviving AB pairs are still in a singlet state at time $t$ is (see Supporting Information, Section S2):

$$p_{\mathrm{AB}}^{\mathrm{S}}(t) = \frac{\mathrm{Tr}\left[\hat{P}_{\mathrm{AB}}^{\mathrm{S}}\hat{\rho}(t)\right]}{\mathrm{Tr}\,[\hat{\rho}(t)]} = \left(\frac{J}{2J+1}\exp\left(-\tfrac{1}{2}k_{\mathrm{C}}^{-}t\right) + \frac{J+1}{2J+1}\exp\left(-\tfrac{1}{2}k_{\mathrm{C}}^{+}t\right)\right)^2 \Big/ s(t), \tag{8}$$

so that in the limit of exclusive formation of the $S = J - ½$ scavenging product:

$$p_{\mathrm{AB}}^{\mathrm{S}}(t \to \infty)|_{k_{\mathrm{C}}^{+}=0} = \frac{J+1}{2J+1}. \tag{9}$$

It is clear from equations (8) and (9) that the scavenging reaction has converted a portion of the surviving radical pairs into the triplet state. When C is a doublet ($J = ½$), 3/4 of the AB pairs survive at $t \to \infty$ and 3/4 of these survivors are singlets (in the AB-manifold). Similarly, if C is a triplet ($J = 1$), 2/3 of the radical pairs survive and 2/3 of them are singlets. If C is a singlet ($J = 0$, i.e. not paramagnetic) there are no spin restrictions on the reaction, all A radicals react with C and there are no radical pairs left at $t \to \infty$. Figure 3 shows $s(t)$ and $p_{\mathrm{AB}}^{\mathrm{S}}(t)$ for scavengers with different spin quantum numbers, $J$.

Additional insight into the origin of singlet-triplet interconversion in AB as a result of the spin-selective AC scavenging reaction is presented in the Appendix (Supporting Information).

**Simulation methods.** The cryptochrome photocycle in Fig. 2a has been modified to include a scavenging reaction (Fig. 2b). We restrict the discussion to a scavenger with spin $J = ½$; preliminary calculations show that qualitatively similar effects can be expected when $J > ½$. The states of the protein, **G** (ground state), **RP** (magnetically sensitive radical pair) and **S** (signalling state) are unchanged. To these is added a fourth state, **X**, formed by a spin-selective reaction of the $\mathrm{FAD}^{\bullet -}$ radical in the **RP** state with a scavenger radical $\mathrm{C}^{\bullet}$. From equation (2), with $J = ½$, **X** comprises the fully oxidised, diamagnetic flavin molecule and the tryptophan radical. The other product of the scavenging reaction is $\mathrm{C}^{-}$, a diamagnetic form of the scavenger. Both **S** and **X** eventually return to the ground state, **G**. Although the primary radical pair in Fig. 2b is shown as containing $\mathrm{WH}^{\bullet +}$, this radical could be further removed from $\mathrm{FAD}^{\bullet -}$ than is the terminal tryptophan of the triad/tetrad in cryptochrome and it need not be a tryptophan radical. In principle, the scavenging reaction could also involve this radical instead of the $\mathrm{FAD}^{\bullet -}$.

The key quantities we wish to calculate are the quantum yields of the reaction intermediates, **S** and **X**, as a function of the direction of an Earth-strength magnetic field (50 μT). The complete equation of motion for the spin density operator is:

$$\frac{\mathrm{d}}{\mathrm{d}t}\hat{\rho}(t, \Omega) = -\mathrm{i}[\hat{H}(\Omega), \hat{\rho}(t, \Omega)] + \hat{\hat{K}}\hat{\rho}(t, \Omega). \tag{10}$$

Here, $\hat{H}(\Omega)$ is the spin Hamiltonian which is the sum of the individual spin Hamiltonians, $\hat{H}_{\mathrm{K}}(\Omega)$, of the three paramagnetic molecules A ($= \mathrm{FAD}^{\bullet -}$), B ($= \mathrm{WH}^{\bullet +}$) and $\mathrm{C}^{\bullet}$, and their electron-electron exchange and dipolar couplings. The $\hat{H}_{\mathrm{K}}(\Omega)$ operators contain the Zeeman interactions with the external magnetic field and hyperfine interactions with surrounding nuclear spins (in angular frequency units):

$$\hat{H}_{\mathrm{K}}(\Omega) = -\gamma_{\mathrm{e}}\mathbf{B}_0(\Omega)\cdot\hat{\mathbf{S}}_{\mathrm{K}} + \sum_{j}^{N_{\mathrm{K}}}\hat{\mathbf{S}}_{\mathrm{K}}\cdot\mathbf{A}_{\mathrm{K}j}\cdot\hat{\mathbf{I}}_{\mathrm{K}j}. \tag{11}$$

$\hat{\mathbf{I}}_{\mathrm{K}j}$ and $\mathrm{A}_{\mathrm{K}j}$ are, respectively, the angular momentum operator and the hyperfine tensor of nuclear spin $j$ in radical K. The sum runs over all $N_{\mathrm{K}}$ magnetic nuclei in radical K. Ω denotes the polar and azimuthal angles specifying the direction of the field in the coordinate frame of the protein. The Larmor frequency, $\nu_0$, is related to the strength of the external magnetic field by $2\pi\nu_0 = |\gamma_{\mathrm{e}}\mathbf{B}_0(\Omega)|$. When $|\mathbf{B}_0(\Omega)| = 50\,\mu\mathrm{T}$, $\nu_0 = 1.4\,\mathrm{MHz}$. The geomagnetic field is weak enough that, for organic radicals, the differences in $g$-values of the three electron spins can safely be neglected.

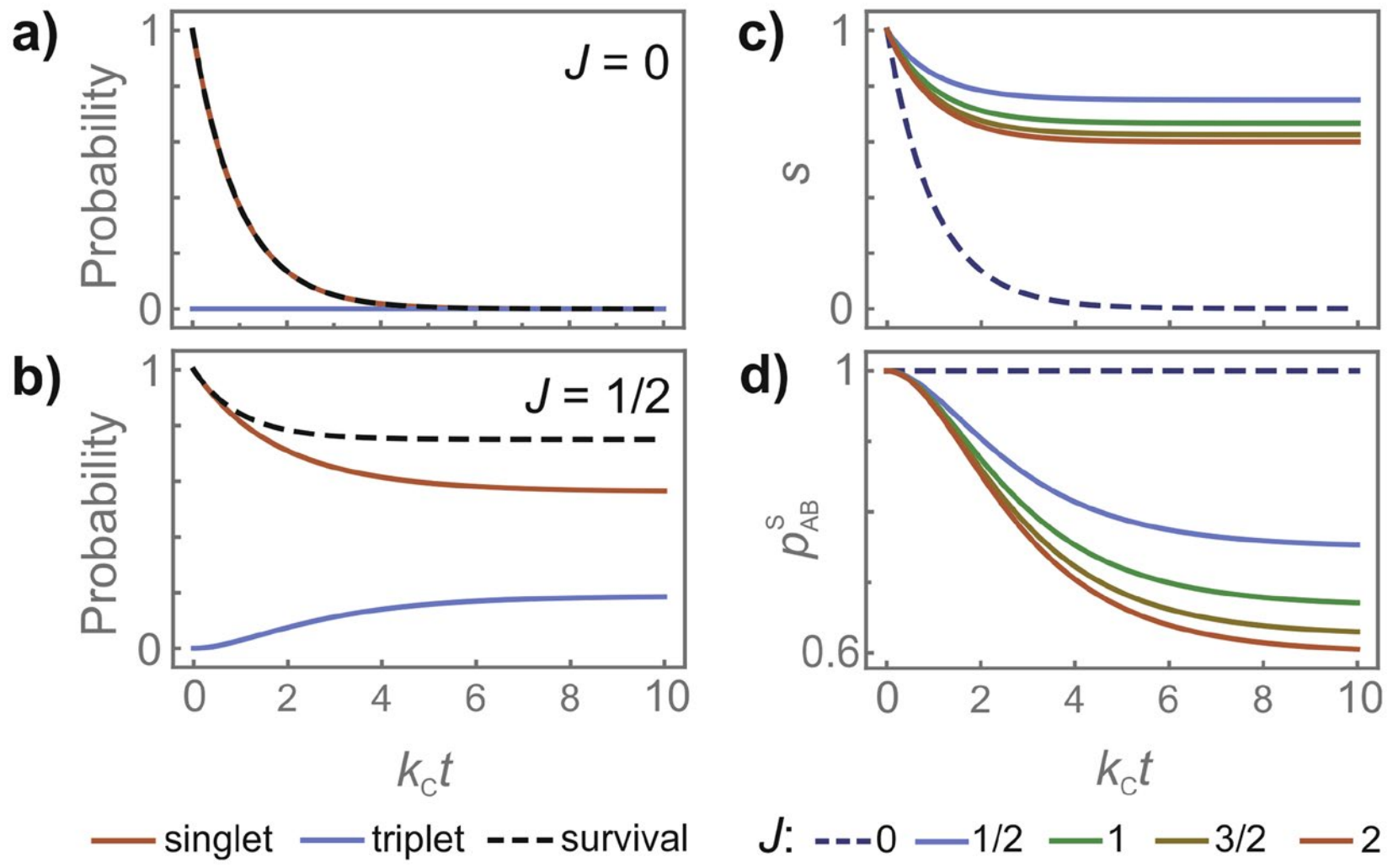

**Figure 3.** Effects of spin-selective scavenging reactions on a model radical pair. Comparison of the singlet fraction ($\mathrm{Tr}[\hat{P}^{S}_{AB}\,\hat{\rho}(t)]$, red lines), the triplet fraction ($\mathrm{Tr}[\,\hat{\rho}(t)] - \mathrm{Tr}[\hat{P}^{S}_{AB}\,\hat{\rho}(t)]$, blue lines) and the survival probability ($s(t)$, dashed black lines) of a radical pair AB reacting with (**a**) a diamagnetic scavenger ($J=0$) and (**b**) a radical scavenger ($J=½$). (**c**) Survival probability, $s(t)$, and (**d**) singlet probability of the survivors, $p^{S}_{AB}(t)$, for a radical pair reacting with scavengers with different spin quantum numbers, $J$. The reaction product has spin quantum number $S=J-½$ (i.e. $k_C^{+}=0$ and $k_C^{-}=k_C$). Additional reaction pathways, coherent spin evolution processes and spin relaxation have all been omitted.

™The reactions of the radical pair are accounted for by the superoperator $\hat{\hat{K}}$ in equation (10). Specifically, $\hat{\hat{K}}$ describes the spin-selective scavenging reaction that forms **X** (equation (4)), the spin-independent formation of **S**, and the charge recombination reaction of the singlet configuration of A and B:

$$\hat{\hat{K}}\hat{\rho}(t,\Omega) = -\frac{1}{2}k_C[\hat{P}^{-}_{AC},\hat{\rho}(t,\Omega)]_{+} - k_f\hat{\rho}(t,\Omega) - \frac{1}{2}k_b[\hat{P}^{S}_{AB},\hat{\rho}(t,\Omega)]_{+}. \tag{12}$$

As in the previous section, we ignore the scavenging reaction that would produce an excited triplet state of $C^{-}$ (equation (1)). $k_C$ is independent of the spin interactions which are all much smaller than the thermal energy $k_BT$. For the singlet initial condition in equation (5), the yield of **S**, once all radicals have reacted, is given by

$$Y_S(\Omega) = k_f\int_0^{\infty}\mathrm{Tr}[\hat{\rho}(t,\Omega)]\mathrm{d}t. \tag{13}$$

We define two quantities as measures of the performance of the radical pair as a magnetic direction sensor: the absolute ($\Delta_S$) and the relative ($\Gamma_S$) anisotropy of the yield of the signalling state **S**:

$$\Delta_S = \max_{\Omega}[Y_S(\Omega)] - \min_{\Omega}[Y_S(\Omega)], \tag{14}$$

$$\Gamma_S = \frac{\Delta_S}{\langle Y_S\rangle} \quad \text{where} \quad \langle Y_S\rangle = \frac{1}{4\pi}\int Y_S(\Omega)d\Omega. \tag{15}$$

It is not clear which of $\Delta_S$ and $\Gamma_S$ corresponds more closely to what the birds perceive when they take a magnetic compass bearing. We therefore present calculations of both in the following. The corresponding quantities, $\Delta_X$ and $\Gamma_X$, for the yield of the scavenging product **X** can be calculated in a similar fashion (see Supporting Information, Figs S1–S4).

## Results

In the following, we explore the effect of scavenging reactions on $\Delta_S$ and $\Gamma_S$ using spin systems of progressively increasing complexity to mimic important aspects of the [$FAD^{\bullet-}$ $WH^{\bullet+}$] radical pair. In every case, the hyperfine tensors and the relative orientation of the radicals were taken from ref. 11. The forward rate constant $k_f$ was fixed at 0.1 $\mu s^{-1}$, consistent with a recent study of spin relaxation in *At*Cry, which suggested that magnetic field effects would be strongly attenuated if the radical pair lifetime exceeded ~10 μs[47]. The charge recombination rate constant, $k_b$, was varied in the range (0, $10k_f$) and its value reported as $\phi = k_b/k_f$ (for purified *At*Cry[23], $\phi$ was found to be ~2). The external magnetic field was 50 μT and spin relaxation was neglected. Unless stated otherwise, the scavenger reacts with the $FAD^{\bullet-}$ radical.

Our starting point is a model with just three $^{14}N$ hyperfine interactions: the N5 and N10 nitrogens in $FAD^{\bullet-}$ and the N1 nitrogen in $WH^{\bullet+}$. We begin with a diamagnetic scavenger ($J=0$) to provide a basis for comparison with the more interesting $J=½$ case. Figure 4a,b show the dependence of the anisotropic yield of **S** on the scavenging rate constant, $k_C$, for several values of $\phi$. The maximum relative anisotropy, $\Gamma_S$, (14.7%) is found when $k_C=0$ (no scavenging) and for $k_b$ at the upper end of the range studied ($\phi=10$, Fig. 4a). $\Gamma_S$ decreases as $\phi$ is reduced, and vanishes for $\phi=0$, when charge recombination (**RP** → **G**) ceases to compete with the forward reaction (**RP** → **S**). Magnetic field effects are expected to be small when the radicals are so far apart that the direct conversion of **RP** to **G** is slow compared to typical spin relaxation times (for a 10 μs relaxation time, this distance is $>\sim$1.7 nm[48]). Both $\Gamma_S$ and $\Delta_S$ are strongly attenuated by fast scavenging ($k_C > 2\pi\nu_0 \approx 10\ \mu s^{-1}$), which allows insufficient time for the 50 μT magnetic field to affect the spin dynamics. The maximum $\Delta_S$ (Fig. 4b) is observed in the absence of scavenging and for intermediate values of $\phi$ (e.g. $\Delta_S=0.055$ when $\phi=2$).

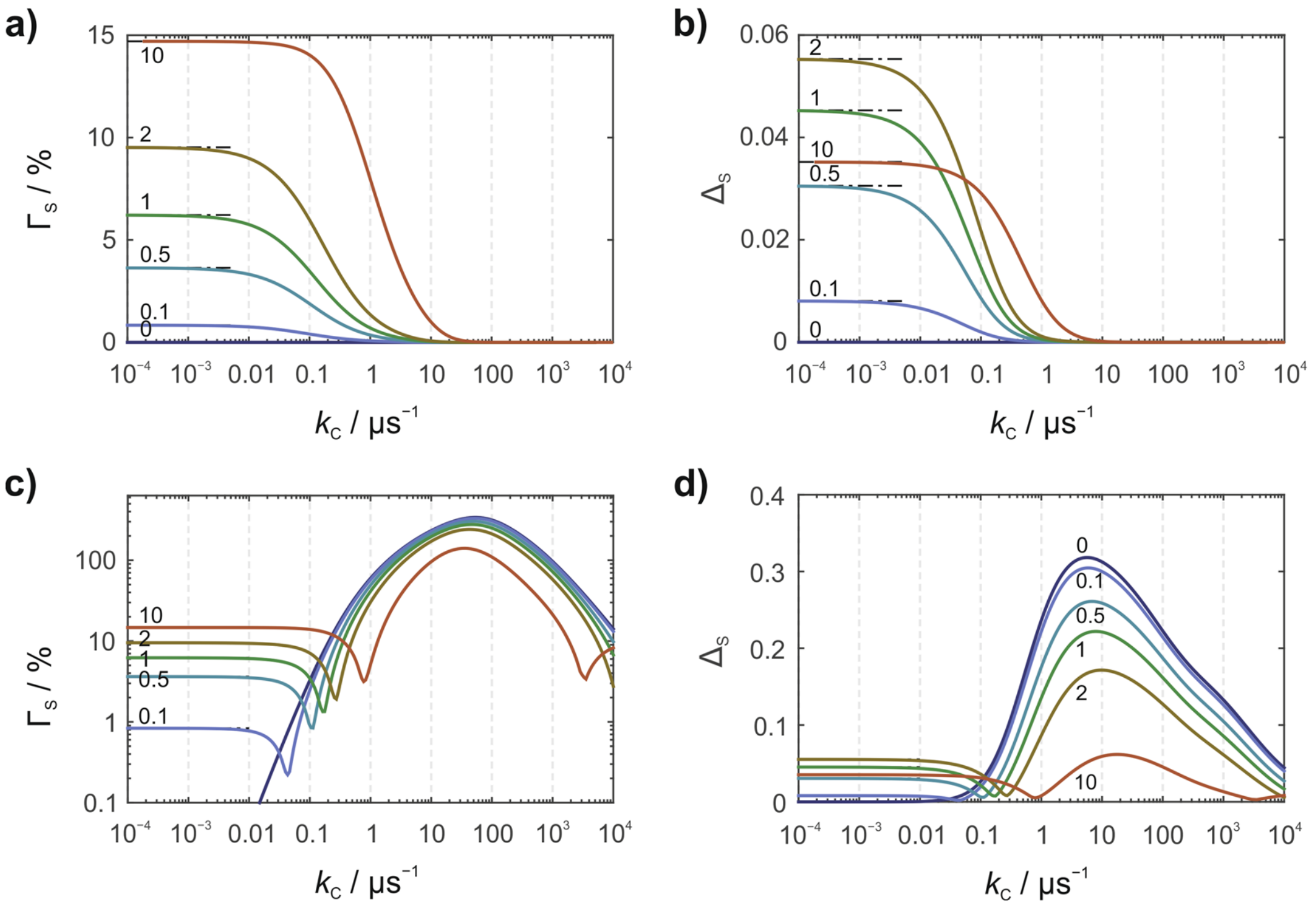

**Figure 4.** Anisotropic yields of the signalling state, **S**, for a model [FAD$^{\bullet-}$ WH$^{\bullet+}$] radical pair. (**a**) and (**c**) relative anisotropy ($\Gamma_S$), (**b**) and (**d**) absolute anisotropy ($\Delta_S$), both as a function of the scavenging rate constant, $k_C$, for various values of $\phi$. The spin system comprises N5 and N10 in FAD$^{\bullet-}$ and N1 in WH$^{\bullet+}$. In (**a**) and (**b**) the scavenger is diamagnetic ($J=0$); in (**c**) and (**d**) it is a radical ($J=½$) with no hyperfine interactions. The product of the scavenging reaction is either a radical (when $J=0$) or a diamagnetic species (when $J=½$).

The situation changes dramatically when the scavenger is paramagnetic ($J=½$). Figure 4c,d show the behaviour of $\Gamma_S$ and $\Delta_S$ for the [FAD$^{\bullet-}$ WH$^{\bullet+}$] model under the same conditions as Fig. 4a,b (but note the logarithmic scale of the vertical axis in Fig. 4c). In this case, C$^{\bullet}$ has a 50 μT electron Zeeman interaction but no hyperfine interactions; more complex systems will be discussed below. The general features of Fig. 4c,d are as follows. (1) Scavenging rate constants well in excess of 10 $\mu s^{-1}$ no longer abolish the magnetic field effect on the signalling state. This is a direct consequence of the unreactivity of the coupled high-spin state ($S=1$) of the flavin radical and the scavenger radical. (2) Large values of $k_C$ lead to significant magnetic field effects even when $k_b$ is small or zero. The scavenging reaction can therefore take over the role played by spin-selective recombination in the current model (Fig. 2a). (3) The scavenging reaction amplifies both $\Gamma_S$ and $\Delta_S$. For this simple model, relative anisotropies, $\Gamma_S$, in excess of 100% are seen for scavenging rate constants in the range 1 to 1000 $\mu s^{-1}$ (Fig. 4c). For $k_b=0$, the maximum $\Gamma_S$ (338%) occurs when $k_C=52\ \mu s^{-1}$. While it is true that these huge anisotropies are accompanied by smaller mean reaction yields, $\langle Y_S \rangle$, the maximum absolute anisotropies, $\Delta_S$ (Fig. 4d), are still larger than those found with a diamagnetic scavenger or with no scavenger at all (Fig. 4b). Qualitatively similar effects under otherwise identical conditions are found when the scavenged radical is WH$^{\bullet+}$ rather than FAD$^{\bullet-}$ (see Supporting Information, Figs S1 and S4).

In view of the enhanced performance of this simple model system, it is important to know whether similar effects can be expected for other radicals. Figure 5 shows the dependence of $\Gamma_S$ and $\Delta_S$ on $k_C$ and $\phi$ for a radical pair, [FAD$^{\bullet-}$ Z$^{\bullet}$], in which WH$^{\bullet+}$ has been replaced by Z$^{\bullet}$, a radical with no hyperfine interactions. Radical pairs of this type have been invoked to explain the reported disorientation of migratory birds exposed to Larmor-frequency magnetic fields[49] and are expected to produce larger magnetic field effects than pairs in which both radicals have significant hyperfine interactions, e.g. [FAD$^{\bullet-}$ WH$^{\bullet+}$][41]. Calculations were performed with N5, N10, H6, 3×H8α and one of the Hβ protons in FAD$^{\bullet-}$ and no hyperfine interactions in the paramagnetic scavenger C. The results (Fig. 5a,b) are qualitatively and quantitatively similar to those in Fig. 4c,d.

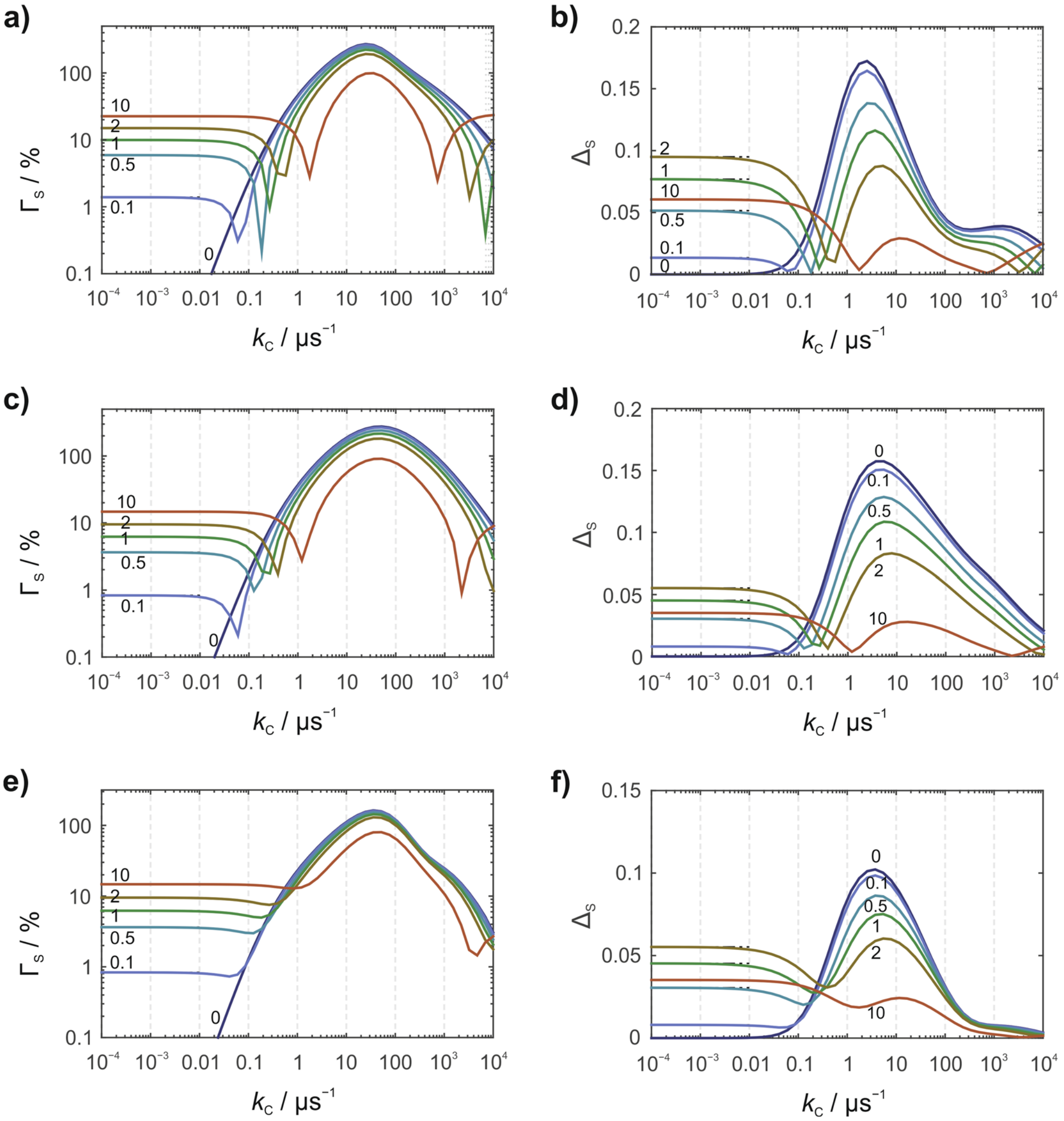

**Figure 5.** Anisotropic yields of the signalling state, **S**, for various model radical pairs. (**a**) and (**b**) [$FAD^{\bullet-}$ $Z^{\bullet}$] radical pair with N5, N10, H6, H8 and Hβ in $FAD^{\bullet-}$ and no hyperfine interactions in $Z^{\bullet}$ or the scavenger. (**c**) and (**d**) [$FAD^{\bullet-}$ $WH^{\bullet+}$] radical pair with N5 and N10 in $FAD^{\bullet-}$ and N1 in $WH^{\bullet+}$. The scavenger had a single isotropic $^{1}$H hyperfine interaction equal to that of the H4 proton in the ascorbyl anion radical[50]. (**e**) and (**f**) [$FAD^{\bullet-}$ $WH^{\bullet+}$] radical pair with N5 and N10 in $FAD^{\bullet-}$ and N1 in $WH^{\bullet+}$. The scavenger, which reacted with $WH^{\bullet+}$, was modelled on $FAD^{\bullet-}$ and included the N5 and N10 hyperfine interactions.

The next step was to test the sensitivity of $\Gamma_S$ and $\Delta_S$ to the presence of hyperfine interactions in the scavenging radical, $C^{\bullet}$. To do this we constructed two additional models using the 3-nucleus version of [$FAD^{\bullet-}$ $WH^{\bullet+}$] above (N5, N10 in $FAD^{\bullet-}$, N1 in $WH^{\bullet+}$). In the first, the scavenger was modelled on a freely diffusing ascorbic acid radical, which is characterised by single dominant $^{1}$H hyperfine coupling[50] and is known to be capable of oxidizing $FAD^{\bullet-}$. In the second, we assumed that the $WH^{\bullet+}$ radical is reduced by a second, uncorrelated $FAD^{\bullet-}$, which was again modelled by means of the N5 and N10 hyperfine interactions. The relative orientations of $WH^{\bullet+}$ and the $FAD^{\bullet-}$ scavenger were chosen (arbitrarily) as those in the crystal structure of dimeric *Dm*Cry[35,36]. In both cases (Fig. 5c–f), the yield of the signalling state showed, once again, remarkably enhanced sensitivity to the direction of the external magnetic field.

Finally, we return to [$FAD^{\bullet-}$ $WH^{\bullet+}$] but now with several hyperfine interactions in both radicals. Figure 6a shows the dependence of $Y_S(\theta)$, the yield of the signalling state, on the direction of the field ($\theta$) in the *yz*-plane of the flavin, for a model comprising N5, N10 and H6 in $FAD^{\bullet-}$ and N1, H1, H4, Hβ and H7 in $WH^{\bullet+}$. In the absence of a spin-selective scavenging reaction (Fig. 6a, red line), there is a "spike" centred at $\theta = 90°$ arising from avoided crossings of spin states with different singlet character. It has been suggested that such a spike could afford a much more precise compass bearing than a smoother, more gently varying $Y_S(\theta)$[11] (see also ref. 34). In the presence of a scavenging reaction (Fig. 6a, black line), a much stronger spike is seen when the field is parallel to the *z*-axis of the flavin ($\theta = 0$; note the very different scales of the vertical axes). Figure 6b shows the dependence of $\Gamma_S$ on the scavenging rate constant. For $\phi = 0$ and $\phi = 2$, the anisotropy is maximized when $k_C = 24\ \mu s^{-1}$ giving $\Gamma_S = 431\%$ and $\Gamma_S = 338\%$, respectively. These figures correspond to 320- and 400-fold increases relative to the case when $k_C = 0$ and $\phi = 2$. The absolute change in the yield of the signalling state is also much larger; for $\phi = 0$, the maximum $\Delta_S$ is 0.248 when $k_C = 3.1\ \mu s^{-1}$.

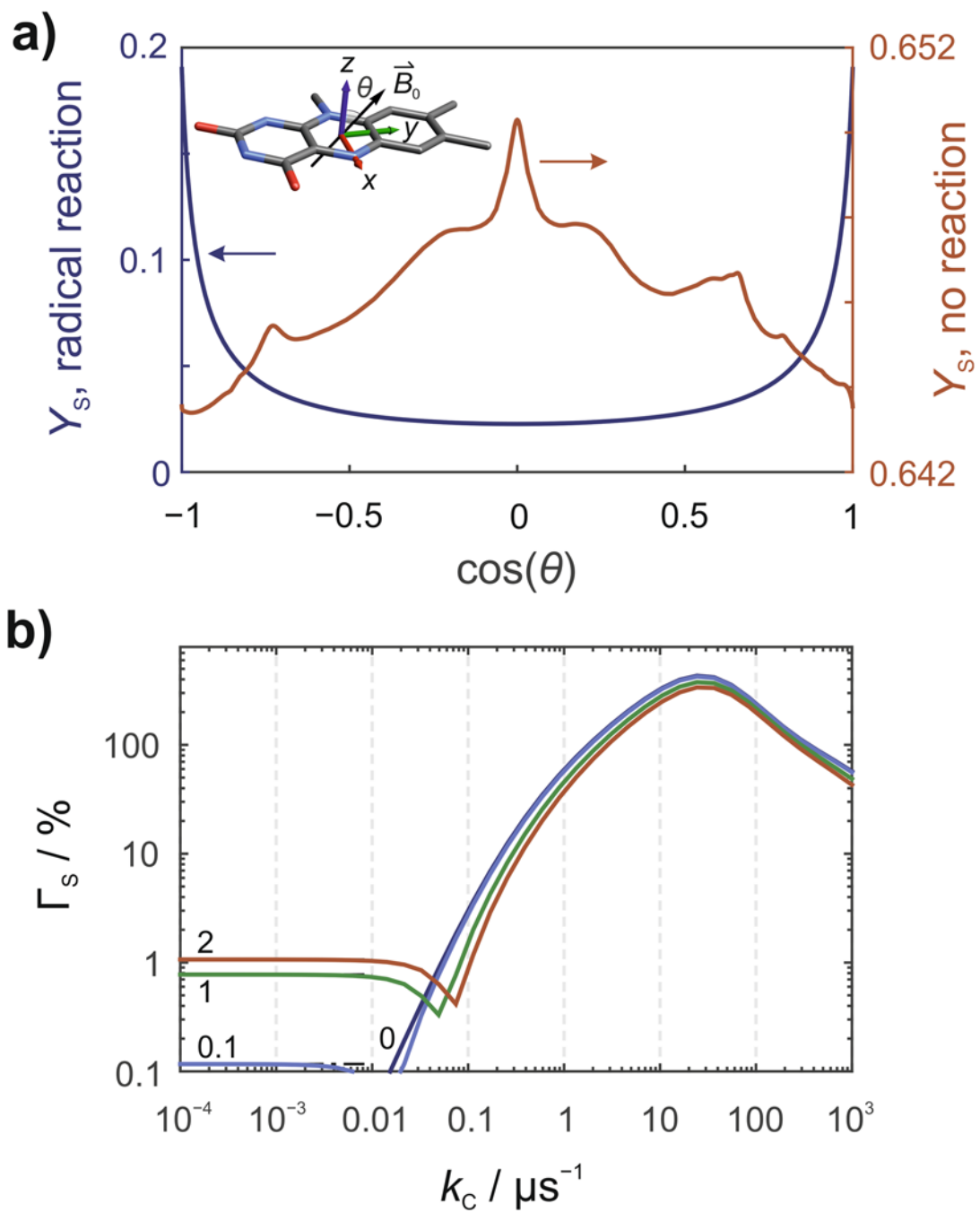

**Figure 6.** Anisotropic yields of the signalling state, **S**, for a model [FAD$^{\bullet-}$ WH$^{\bullet+}$] radical pair. (**a**) The yield of **S**, $Y_S$, as a function of the direction of the magnetic field in the *yz*-plane of the flavin for $k_C = 0$, $k_b = 2k_f$ (red line) and $k_C = 24.2\ \mu s^{-1}$, $k_b = 0$ (blue line). Note the very different vertical scales for the two traces. (**b**) The relative yield anisotropy, $\Gamma_S$, as a function of $k_C$ and $\phi$. The spin system comprised N5, N10 and H6 in FAD$^{\bullet-}$ and N1, H1, H4, H$\beta$ and H7 in WH$^{\bullet+}$. The scavenger had no hyperfine interactions.

A tentative explanation of the large anisotropy of the yield of the signalling state is presented in the Appendix (Supporting Information).

We have focussed here on the effects of scavenging reactions on the anisotropic yield of the state **S**. Relative and absolute anisotropies have also been calculated for **X**, the product of the scavenging reaction. Broadly speaking, $\Gamma_X$ and $\Delta_X$ show behaviour similar to $\Gamma_S$ and $\Delta_S$ except that the anisotropy has opposite sign (see Supporting Information, Figs S1–S4). $\Gamma_X$ peaks at values of $k_C$ about 10 times smaller than does $\Gamma_S$, with amplitudes of several tens of percent instead of the hundreds of percent found for **S**. In order to benefit from the signal amplification, it is essential therefore that the scavenger should not be capable of converting **S** into **X**. This could be prevented if the radicals in **S** and **RP** have different protonation states (as indicated in Fig. 2b). For the sake of illustration, oxidation of flavin semiquinone radicals by $O_2$ is $10^4$ times faster for the anionic radical than for the neutral radical[51,52]. Clearly the details will depend on the identity and reactivity of the radical pair and the scavenger.

## Discussion

**Amplification.** The notion that magnetoreception could rely on quantum coherence in transient radical pairs has caught the imagination of scientists in a range of disciplines from zoology to theoretical physics[53–58]. Although evidence in support of the hypothesis is accumulating, the sensor molecules have yet to be unequivocally identified[2]. Although cryptochrome seems to be required for a number of magnetic responses in fruit flies[59–68], there is no definite proof yet that cryptochrome functions as the magnetic sensor or that the *Drosophila* findings have a direct bearing on the mechanism of compass magnetoreception in birds. Moreover, it is not clear whether the magnetic field effects observed for purified cryptochromes *in vitro*[23,39,69], or the reaction scheme that accounts for them, are identical to those *in vivo*[2]. It is therefore appropriate to explore alternative photocycles. We have chosen in this work to focus on cryptochrome because, 17 years after it was first proposed[9], it is still the only candidate magnetoreceptor for the avian compass and because FAD radicals seem to be near optimal as components of a direction sensor[41].

A potential problem with the current model based on [FAD$^{\bullet-}$ WH$^{\bullet+}$] (Fig. 2a) is that the predicted anisotropy of the reaction yield is tiny[41]. This is a consequence of the number and lack of symmetry of the hyperfine interactions in WH$^{\bullet+}$ [41] and the inevitable relaxation of the spin-coherence which is expected to attenuate the anisotropy if the radical pairs live for more than a few microseconds[15,47,70,71]. A recent study suggested that the anisotropic magnetic field effect ($\Gamma_S$) would be of the order of 0.1% for a realistic model of [FAD$^{\bullet-}$ WH$^{\bullet+}$] including spin relaxation[47]. Even if evolutionary pressure has somehow solved the problem of decoherence, the effects are still expected to be small. For long-lived, slowly relaxing [FAD$^{\bullet-}$ WH$^{\bullet+}$] radical pairs, it has been proposed that spikes in $Y_S(\Omega)$, predicted for certain directions of the field, could improve the precision of the compass bearing[11] (see also ref. 34). Nevertheless, $\Gamma_S$ is not expected to be much larger than ~1%. A [$FAD^{\bullet-}$ $Z^{\bullet}$] radical pair, with no magnetic nuclei in $Z^{\bullet}$, could give much stronger signals[41], although the only obvious $Z^{\bullet}$ radical (superoxide, $O_2^{\bullet-}$) is almost certainly unsuitable due to its exceedingly fast spin relaxation[72]. Furthermore, spikes in $Y_S(\Omega)$ are not expected for a [$FAD^{\bullet-}$ $Z^{\bullet}$] pair[11] meaning that the compass precision would be inferior to that of a radical pair with hyperfine interactions in both radicals.

In any case, amplification of the primary signal will be an essential feature of the biological compass[34]. This could occur within the sensor itself and/or in the course of signal transduction. A cyclic kinetic scheme for amplification of the primary effect has recently been suggested[40]. It relies on the effects of slow radical termination reactions on the photo-stationary state of the continuously illuminated protein. Although this process has been demonstrated experimentally for a purified cryptochrome[40], it is not clear how well it would work for nocturnally migrating birds whose magnetic compass appears to function at very low light levels[2,73]. The amplification scheme suggested here (Fig. 2b) is practically acyclic and hence not subject to the same limitations. Furthermore, the scavenging process gives rise to remarkably large anisotropy in the reaction yields (more than 100% in all the models considered here), which would drastically reduce (by a factor of $10^4$ to $10^6$) the number of integrated events required to elicit a directional response with the required signal-to-noise ratio[11]. As in the current model[11], there is a spike in the reaction yield which becomes narrower and more pronounced as the lifetime of the radical pair is increased (see Supporting Information, Fig. S6). In contrast to the current model, this feature occurs when the field is parallel ($\theta = 0$) rather than perpendicular ($\theta = 90°$) to the normal to the plane of the flavin moiety. Such spikes offer the possibility of highly precise compass bearings[11,34]. Furthermore, the shape of the reaction yield anisotropy is to a large extent independent of the details of the scavenging process, including the hyperfine interactions in all three radicals, such that similar responses can be expected under a variety of conditions (see Supporting Information, Fig. S5).

**Distance constraints.** A spin-selective reaction channel is required in order that the effect of the magnetic field on the coherent spin motion can alter the yield of the reaction product. In the scheme shown in Fig. 2a, this is the charge recombination step. The rate constant of this reaction, $k_b$, must be larger than or comparable to the electron spin relaxation rate, otherwise the reaction yield will simply reflect the statistical ratio of singlet and triplet states, which is essentially independent of the direction of a weak magnetic field. Assuming this reaction occurs in a single step (rather than by sequential electron hopping), this requirement precludes radical pairs with edge-to-edge distances greater than ~1.7 nm. According to the empirical "Moser-Dutton ruler", electron transfer rate constants as small as $10^3\,s^{-1}$ are expected for an edge-to-edge distance of 2 nm, even in the Marcus activation-less limit[48]. Furthermore, an increase of 0.33 nm in the separation of the radicals is expected to result in a 100-fold decrease in the rate constant of charge recombination. In principle, this restriction on the separation of the radicals could be relaxed if charge recombination occurred indirectly, either by reversible electron hopping along the Trp-triad/tetrad or via a second, independent, electron transfer chain. Given the strongly exergonic nature of the forward electron transfer (reflecting the steadily increasing solvent-exposure of the radical pair as the charge is propagated along the electron transfer chain)[74] neither possibility seems likely.

This distance-constraint has important implications for the operation of a radical pair compass. It appears that animal cryptochromes (including those of birds) generally contain a fourth tryptophan, $Trp_DH$, which extends the tryptophan triad to a tetrad. With an edge-to-edge distance of ~1.7 nm, no magnetic field effects would be expected for [$FAD^{\bullet-}$ $Trp_DH^{\bullet+}$] in the current model (Fig. 2a) if the radical lifetime is of the order of several microseconds. Recent studies of *Dm*Cry[38,39] and *Xenopus laevis* (6–4) photolyase[75] suggest that this is indeed the case. Homology modelling of robin (*Erithacus rubecula*) Cry1a predicts an edge-to-edge distance of 1.96 nm for [$FAD^{\bullet-}$ $Trp_DH^{\bullet+}$]. The absence of spin-selective recombination of these distant radical pairs on a microsecond timescale calls into question the current model (Fig. 2a). Our new scheme (Fig. 2b) not only liberates the model from the spatial constraints imposed by the charge recombination step, but also strongly amplifies the magnetic field effect. The scheme is even applicable when the partner of the $FAD^{\bullet-}$ radical is free to diffuse rather than attached to the cryptochrome. If such a small, rapidly tumbling radical reacted spin-selectively with a freely diffusing paramagnetic scavenger, there would be the added benefit that both would undergo slower spin relaxation than the protein-bound radicals. Modulation of the exchange interactions of freely diffusing paramagnetic species may cause singlet-triplet dephasing in the radical pair which could either attenuate or further amplify the anisotropy[15].

Finally, lifting the restriction on the rate of the spin-selective recombination reaction also means that the detrimental effects of inter-radical exchange and dipolar interactions[2,76] can be minimised by placing the radicals much further apart than would be permissible in the current model.

**Properties of the radicals.** In all the cases analysed above, the enhanced yields of the signalling state are largely independent of the hyperfine interactions in both the paramagnetic scavenger and $WH^{\bullet+}$. In particular, the reduction in the magnetic field effect caused by the hyperfine interactions in $WH^{\bullet+}$ for the current model is not found for the scavenging reaction scheme. The mechanism does not require particular properties of the scavenger except that its spin relaxation is slow on the timescale of the spin dynamics and the recombination reactions. In practice, this suggests that the mechanism is feasible if spin relaxation in the scavenger occurs at a rate comparable to that in the primary radical pair. This restriction probably excludes certain rapidly relaxing species such as superoxide[72] and many transition metal complexes. Disregarding this aspect for the moment, iron-sulphur clusters could in principle act as scavengers. We mention this in the light of the recent report of a multimeric complex of cryptochrome and a protein, IscA, containing [2Fe-2S] clusters[77]. While several aspects of this work are controversial, not least the claim that the complex possesses a permanent magnetic moment[78], the reported structure is potentially interesting in the context of the mechanism suggested here. However, the tryptophan triad/tetrad in cryptochrome is probably too far away from any of the [2Fe-2 S] clusters for a sufficiently rapid electron transfer reaction.

In principle, molecules with electron spin greater than ½ (e.g. $O_2$, $J = 1$) could act as scavengers. Although this again raises the question of rapid spin relaxation[72], it opens the interesting prospect of directly linking the spin dynamics in the cryptochrome radical pair to redox-active ion channels, which have been implicated in neuronal firing in fruit flies[66, 79].

## Conclusions

We have demonstrated that the directional response, both relative and absolute, of a radical pair to the Earth's magnetic field can be significantly enhanced when one of the radicals can react with an external paramagnetic molecule. This scavenging reaction acts as a spin-selective recombination channel resulting in a field-dependent product yield even when spin-selective charge recombination in the radical pair is very slow. As a consequence, the radical pair mechanism is freed from the constraint that the constituents of the radical pair must be less than about 1.7 nm apart (edge-to-edge) in order that direct charge recombination is fast enough to compete with spin relaxation. We believe that our suggestion may have far-reaching implications for the detailed operation of the proposed quantum compass in birds.

## Acknowledgements

This work was supported by the European Research Council (under the European Union's 7th Framework Programme, FP7/2007-2013/ERC grant agreement no. 340451), the Air Force Office of Scientific Research (Air Force Materiel Command, USAF award no. FA9550-14-1-0095), and the EMF Biological Research Trust. We are grateful to Emil Sjulstok and Ilia Solov'yov for sharing their homology model of robin Cry1a (*Erithacus rubecula*) and to Marco Bassetto for helpful suggestions of potential signalling pathways in *D. melanogaster*.

## Author Contributions

D.R.K. conceived the idea behind this work, performed the research, and discussed the results with P.J.H. Both authors wrote the paper.

## Additional Information

**Supplementary information** accompanies this paper at doi:10.1038/s41598-017-09914-7

**Competing Interests:** The authors declare that they have no competing interests.

**Publisher's note:** Springer Nature remains neutral with regard to jurisdictional claims in published maps and institutional affiliations.

# The sensitivity of a radical pair compass magnetoreceptor can be significantly amplified by radical scavengers

Daniel R. Kattnig† and P. J. Hore*

Department of Chemistry, University of Oxford, Physical and Theoretical Chemistry Laboratory, South Parks Road, Oxford OX1 3QZ, U.K.

* Author for correspondence: peter.hore@chem.ox.ac.uk
† Permanent address: Department of Physics, University of Exeter, Living Systems Institute, Stocker Road, Exeter, EX4 4QD

## Appendix

## Supporting Information

# Appendix

## Singlet-triplet interconversion in AB

Insight into the origin of singlet-triplet interconversion in AB as a result of a spin-selective AC scavenging reaction may be obtained from the following simple argument.[36]

First, we define the usual singlet and triplet states for the AB and AC pairs:

$$\begin{aligned} \left|\mathrm{S}^{\mathrm{AX}}\right\rangle &= \frac{1}{\sqrt{2}}\left|\alpha^{\mathrm{A}}\beta^{\mathrm{X}}\right\rangle - \frac{1}{\sqrt{2}}\left|\beta^{\mathrm{A}}\alpha^{\mathrm{X}}\right\rangle \\ \left|\mathrm{T}_{+1}^{\mathrm{AX}}\right\rangle &= \left|\alpha^{\mathrm{A}}\alpha^{\mathrm{X}}\right\rangle \\ \left|\mathrm{T}_{0}^{\mathrm{AX}}\right\rangle &= \frac{1}{\sqrt{2}}\left|\alpha^{\mathrm{A}}\beta^{\mathrm{X}}\right\rangle + \frac{1}{\sqrt{2}}\left|\beta^{\mathrm{A}}\alpha^{\mathrm{X}}\right\rangle \\ \left|\mathrm{T}_{-1}^{\mathrm{AX}}\right\rangle &= \left|\beta^{\mathrm{A}}\beta^{\mathrm{X}}\right\rangle, \end{aligned} \tag{1}$$

where X = B or C. We start with the AB pair in its singlet state, $\left|\mathrm{S}^{\mathrm{AB}}\right\rangle$, and the radical C in state $\left|\alpha^{\mathrm{C}}\right\rangle$ ($\left|\mathrm{S}^{\mathrm{AB}}\beta^{\mathrm{C}}\right\rangle$ is considered below). This initial state, $\left|\psi\right\rangle = \left|\mathrm{S}^{\mathrm{AB}}\alpha^{\mathrm{C}}\right\rangle$ in the $\{\mathrm{AB}\}\mathrm{C}$ basis, can be expressed in the ABC product basis using equations (1):

$$\left|\psi\right\rangle = \frac{1}{\sqrt{2}}\left|\alpha^{\mathrm{A}}\beta^{\mathrm{B}}\alpha^{\mathrm{C}}\right\rangle - \frac{1}{\sqrt{2}}\left|\beta^{\mathrm{A}}\alpha^{\mathrm{B}}\alpha^{\mathrm{C}}\right\rangle, \tag{2}$$

and then transformed into the $\{\mathrm{AC}\}\mathrm{B}$ basis, again using equations (1):

$$\left|\psi\right\rangle = \frac{1}{2}\left|\mathrm{S}^{\mathrm{AC}}\alpha^{\mathrm{B}}\right\rangle - \frac{1}{2}\left|\mathrm{T}_{0}^{\mathrm{AC}}\alpha^{\mathrm{B}}\right\rangle + \frac{1}{\sqrt{2}}\left|\mathrm{T}_{+1}^{\mathrm{AC}}\beta^{\mathrm{B}}\right\rangle. \tag{3}$$

Equation (3) shows that the AC pair is 25% singlet and 75% triplet, as would be expected from the absence of correlation between C and either A or B.

Now we allow the AC singlets to recombine via a spin-selective scavenging reaction. To see the effect most clearly, we simply remove the first term on the right hand side of equation (3) to give the modified state $\left|\psi'\right\rangle$:

$$\left|\psi'\right\rangle = -\frac{1}{2}\left|\mathrm{T}_{0}^{\mathrm{AC}}\alpha^{\mathrm{B}}\right\rangle + \frac{1}{\sqrt{2}}\left|\mathrm{T}_{+1}^{\mathrm{AC}}\beta^{\mathrm{B}}\right\rangle. \tag{4}$$

$\left|\psi'\right\rangle$ can be transformed back into the ABC product basis:

$$\left|\psi'\right\rangle = -\frac{1}{2\sqrt{2}}\left|\alpha^{\mathrm{A}}\alpha^{\mathrm{B}}\beta^{\mathrm{C}}\right\rangle + \frac{1}{\sqrt{2}}\left|\alpha^{\mathrm{A}}\beta^{\mathrm{B}}\alpha^{\mathrm{C}}\right\rangle - \frac{1}{2\sqrt{2}}\left|\beta^{\mathrm{A}}\alpha^{\mathrm{B}}\alpha^{\mathrm{C}}\right\rangle, \tag{5}$$

and then into the original $\{\mathrm{AB}\}\mathrm{C}$ basis:

$$\left|\psi'\right\rangle = \frac{3}{4}\left|\mathrm{S}^{\mathrm{AB}}\alpha^{\mathrm{C}}\right\rangle + \frac{1}{4}\left|\mathrm{T}_{0}^{\mathrm{AB}}\alpha^{\mathrm{C}}\right\rangle - \frac{1}{2\sqrt{2}}\left|\mathrm{T}_{+1}^{\mathrm{AB}}\beta^{\mathrm{C}}\right\rangle. \tag{6}$$

Renormalizing $|\psi'\rangle$ gives:

$$|\bar{\psi}'\rangle = \frac{\sqrt{3}}{2}|S^{AB}\alpha^{C}\rangle + \frac{1}{2\sqrt{3}}|T_0^{AB}\alpha^{C}\rangle - \frac{1}{\sqrt{6}}|T_{+1}^{AB}\beta^{C}\rangle . \quad (7)$$

The proportions of singlet and triplet AB pairs, which were initially 100% and 0% respectively, are now 75% and 25%. If we start with $|\psi\rangle = |S^{AB}\beta^{C}\rangle$ instead of $|S^{AB}\alpha^{C}\rangle$, the equivalent of equation (7) is:

$$|\bar{\psi}'\rangle = \frac{\sqrt{3}}{2}|S^{AB}\beta^{C}\rangle - \frac{1}{2\sqrt{3}}|T_0^{AB}\beta^{C}\rangle + \frac{1}{\sqrt{6}}|T_{-1}^{AB}\alpha^{C}\rangle \quad (8)$$

which again gives 75% singlet and 25% triplet. Thus the net effect of the spin-selective AC reaction is to induce singlet-triplet interconversion in the AB pair even though there is initially no spin correlation between C and either A or B.

## Signalling state anisotropy

The principal factor behind the unexpectedly large values of $\Delta_S$ and $\Gamma_S$ appears to be the form of the hyperfine interactions of the N5 and N10 nitrogens in $FAD^{\bullet-}$. As the most anisotropic hyperfine interactions in the flavin radical, they seem to reinforce one another and to dominate the spin dynamics of $FAD^{\bullet-}$-containing radical pairs.[34] Both $^{14}N$ hyperfine tensors have almost perfect axial symmetry, with parallel symmetry axes, large *z*-components and near-zero *x*- and *y*-components.[11, 34] A consequence is that when the magnetic field is parallel to the symmetry axis, the spin Hamiltonian connects the AB singlet state to $T_0^{AB}$ but not to $T_{+1}^{AB}$ or $T_{-1}^{AB}$. By contrast, when the field is perpendicular to the hyperfine symmetry axis, $S^{AB}$ is mixed with all three AB triplet states.

In the parallel configuration, the $S^{AB} \leftrightarrow T^{AB}$ interconversion caused by the AC scavenging reaction (see the Appendix in the main text) leads to $T^{AB}$ states which (a) cannot be converted to $S^{AB}$ by the spin Hamiltonian and are therefore unable to return to the ground state, (b) are not scavenged because of the Wigner spin-conservation requirements, and which therefore (c) contribute to a high yield of the non-selectively formed signalling state. It is these states that are responsible for the long-time behaviour shown in Fig. 3b (main text).

In the perpendicular case, the more extensive $S^{AB} \leftrightarrow T^{AB}$ mixing means that no $T^{AB}$ states are immune to spin-selective recombination and scavenging. The result is a lower yield of the competing reaction that leads to the signalling state. Intermediate orientations show similar behaviour. It is only when the field is parallel to the dominant hyperfine axis that singlet-triplet mixing in the AB pair becomes restricted and the spike emerges. This qualitative difference between parallel and all other directions of the magnetic field seems to be responsible for the large anisotropies in the yield of the signalling state.

# Figure S1

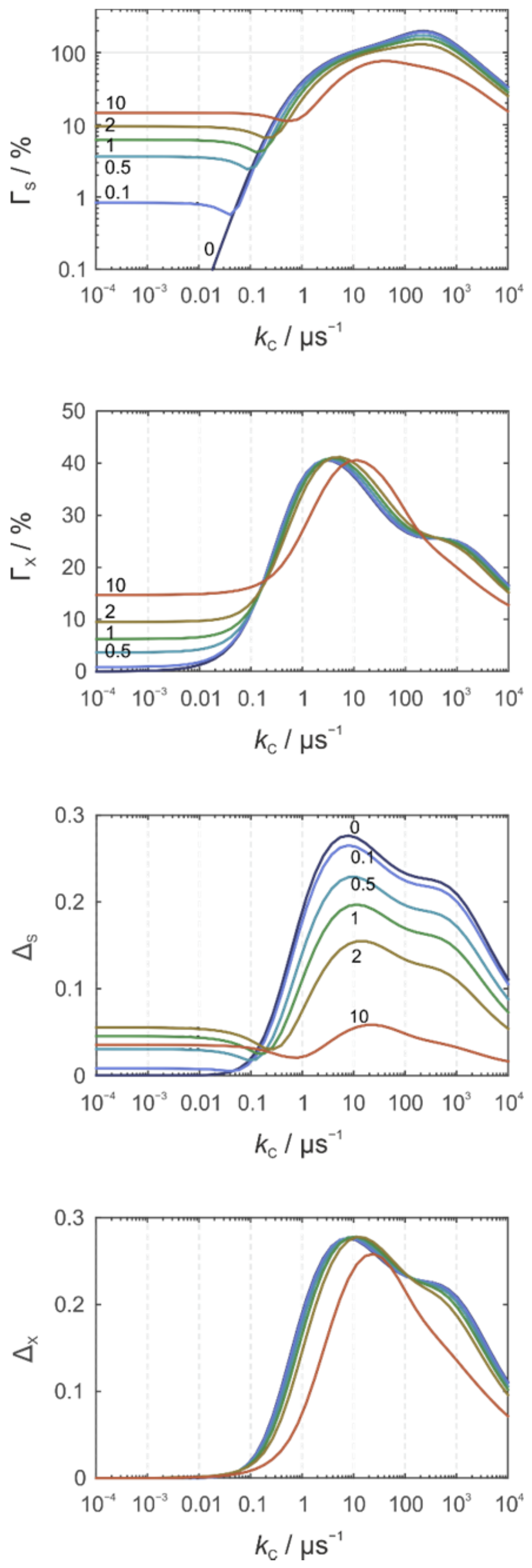

**Anisotropic yields of the signalling state, S, and the scavenging product, X, for a model $[FAD^{\bullet-}\ WH^{\bullet+}]$ radical pair**. The scavenger is a radical ($J$ = ½) with no hyperfine interactions. (a) and (b) relative anisotropies ($\Gamma_S$ and $\Gamma_X$, respectively), (c) and (d) absolute anisotropies ($\Delta_S$ and $\Delta_X$, respectively), both as a function of the scavenging rate constant, $k_C$, for various values of $\phi$. The spin system comprises N5 and N10 in $FAD^{\bullet-}$ and N1 in $WH^{\bullet+}$. The model is identical to that used for Figures 4(c) and 4(d) except that the scavenger reacted with $W^{\bullet+}$ instead of $FAD^{\bullet-}$. $\Gamma_X$ and $\Delta_X$ are defined by analogy with $\Gamma_X$ and $\Delta_X$ (equations (15) and (14), respectively).

## Figure S2

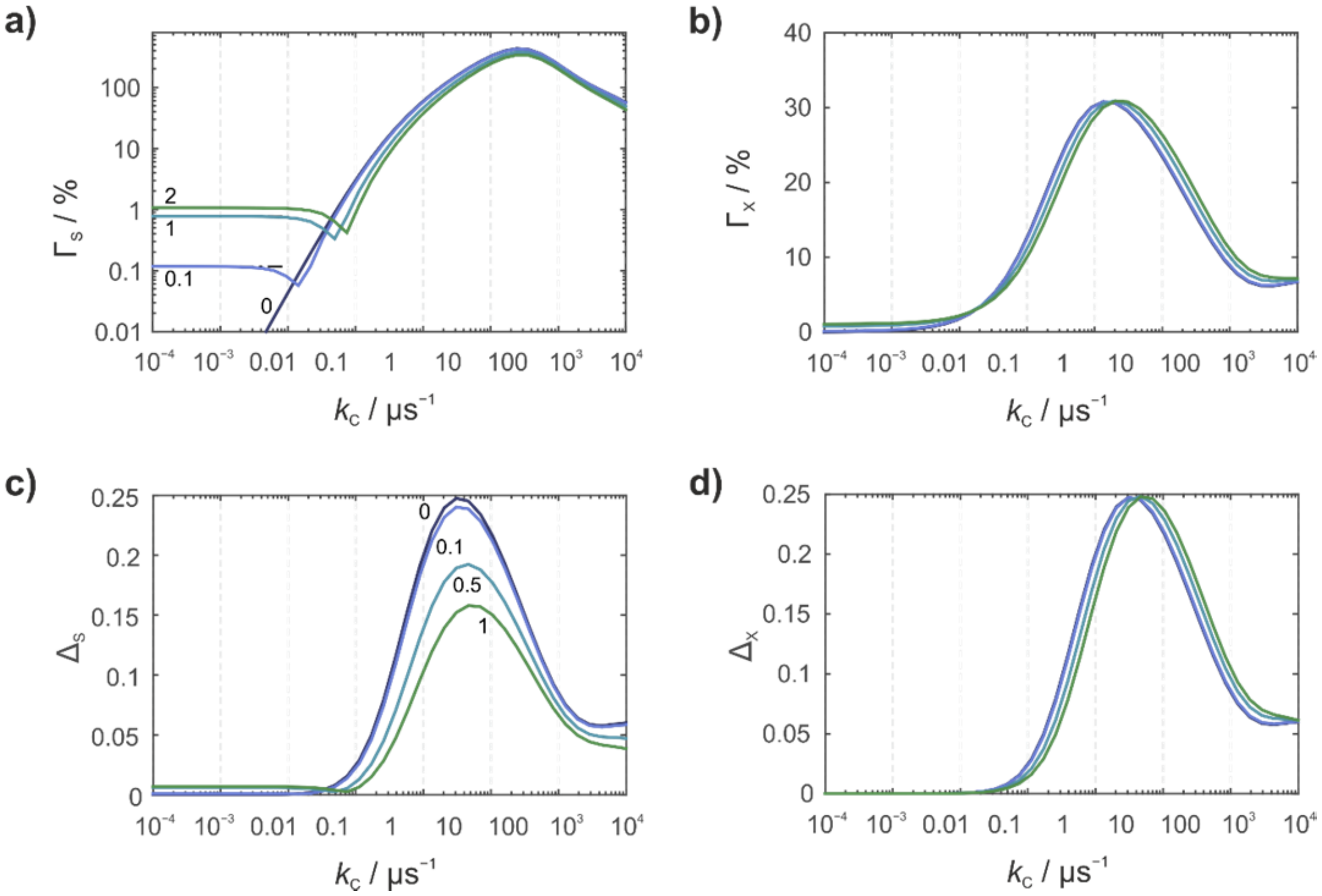

**Anisotropic yields of the signalling state, S, and the scavenging product, X, for a model [FAD$^{•-}$ WH$^{•+}$] radical pair**. The scavenger is a radical ($J$ = ½) with no hyperfine interactions. (a) and (b) relative anisotropies ($\Gamma_S$ and $\Gamma_X$, respectively), (c) and (d) absolute anisotropies ($\Delta_S$ and $\Delta_X$, respectively), both as a function of the scavenging rate constant, $k_C$, for various values of $\phi$. The spin system comprises N5, N10 and H6 in FAD$^{•-}$ and N1, H1, H4, Hβ and H7 in WH$^{•+}$. The scavenger reacted with FAD$^{•-}$.

## Figure S3

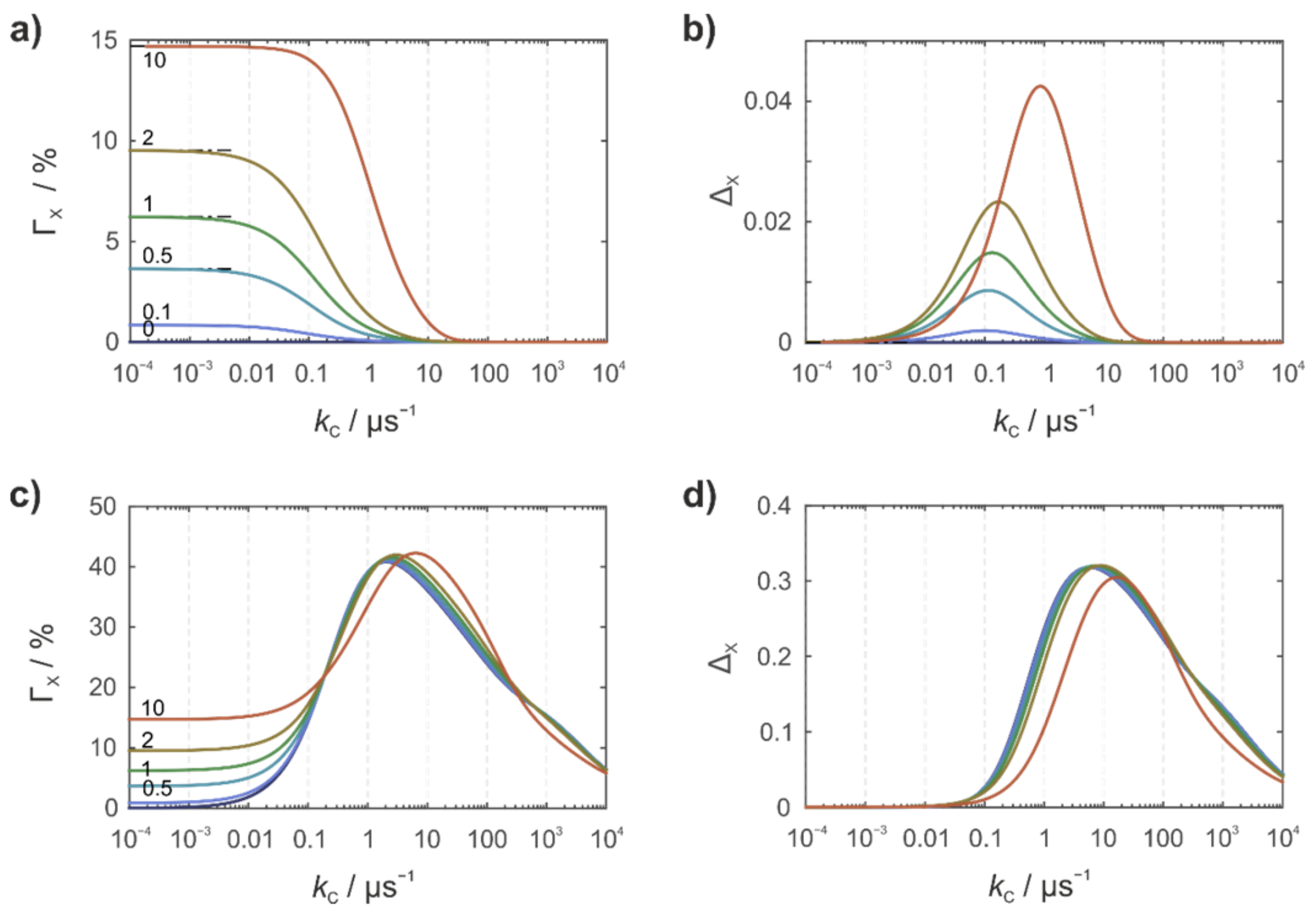

**Anisotropic yields of the scavenging product, X, for a model [FAD$^{\bullet-}$ WH$^{\bullet+}$] radical pair**. (a) and (c) relative anisotropies ($\Gamma_X$), (b) and (d) absolute anisotropies ($\Delta_X$), both as a function of the scavenging rate constant, $k_C$, for various values of $\phi$. In (a) and (b) the scavenger is diamagnetic ($J$ = 0); in (c) and (d) it is a radical ($J$ = ½) with no hyperfine interactions. The spin system comprises N5 and N10 in FAD$^{\bullet-}$ and N1 in WH$^{\bullet+}$. The scavenger reacted with FAD$^{\bullet-}$.

# Figure S4

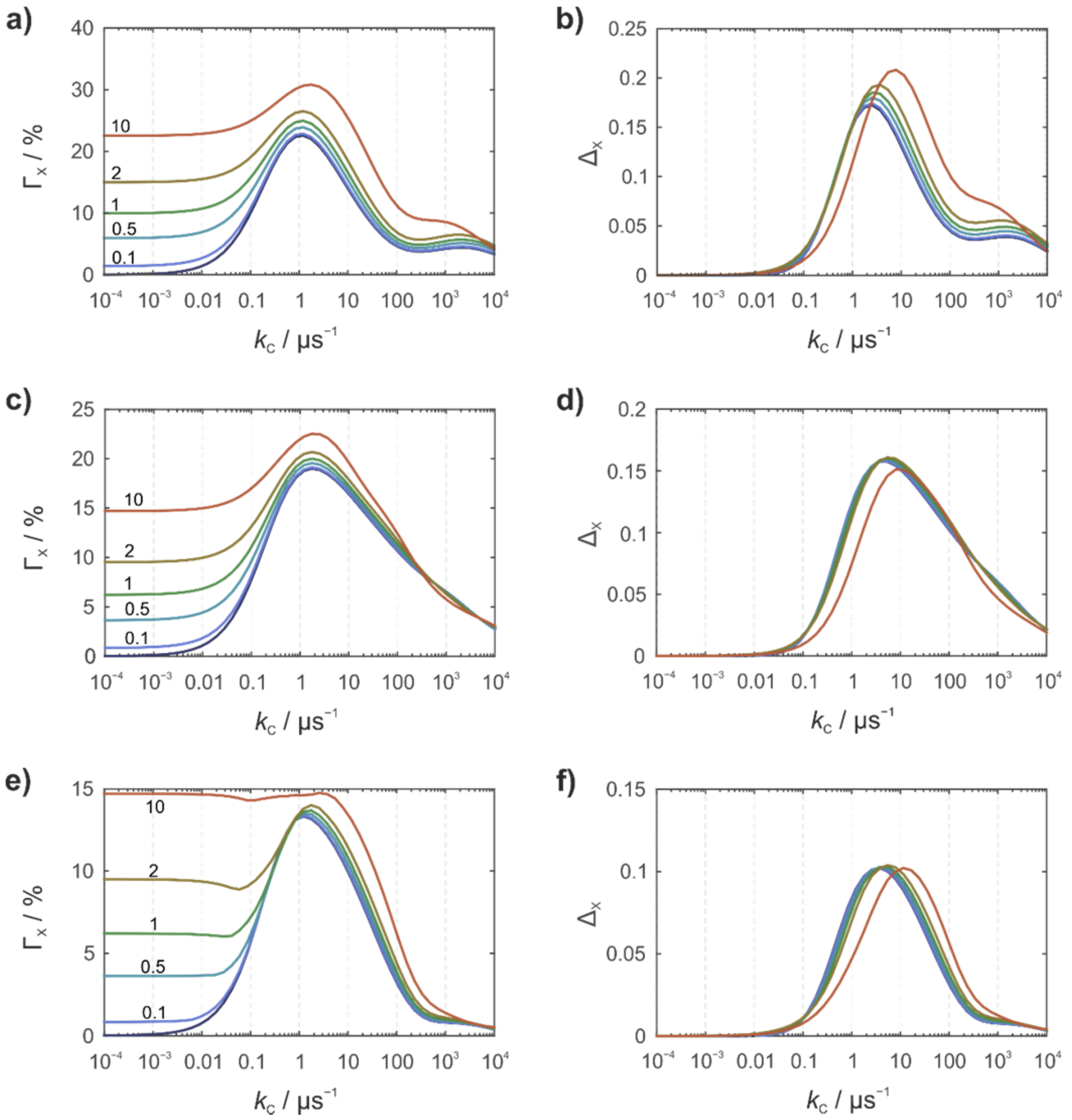

**Anisotropic yields of the scavenging product, X, for various model radical pairs.** (a) and (b) [$FAD^{\bullet-}$ $Z^{\bullet}$] radical pair with N5, N10, H6, H8 and Hβ in $FAD^{\bullet-}$ and no hyperfine interactions in $Z^{\bullet}$ or the scavenger. (c) and (d) [$FAD^{\bullet-}$ $WH^{\bullet+}$] radical pair with N5 and N10 in $FAD^{\bullet-}$ and N1 in $WH^{\bullet+}$. The scavenger had a single isotropic $^{1}H$ hyperfine interaction equal to that of the H4 proton in the ascorbyl anion radical[40]. (e) and (f) [$FAD^{\bullet-}$ $WH^{\bullet+}$] radical pair with N5 and N10 in $FAD^{\bullet-}$ and N1 in $WH^{\bullet+}$. The scavenger, which reacted with $WH^{\bullet+}$, was modelled on $FAD^{\bullet-}$ and included the N5 and N10 hyperfine interactions. These calculations are identical to those used for Figure 5 except that $\Gamma_X$ and $\Delta_X$ are shown instead of $\Gamma_S$ and $\Delta_S$.

# Figure S5

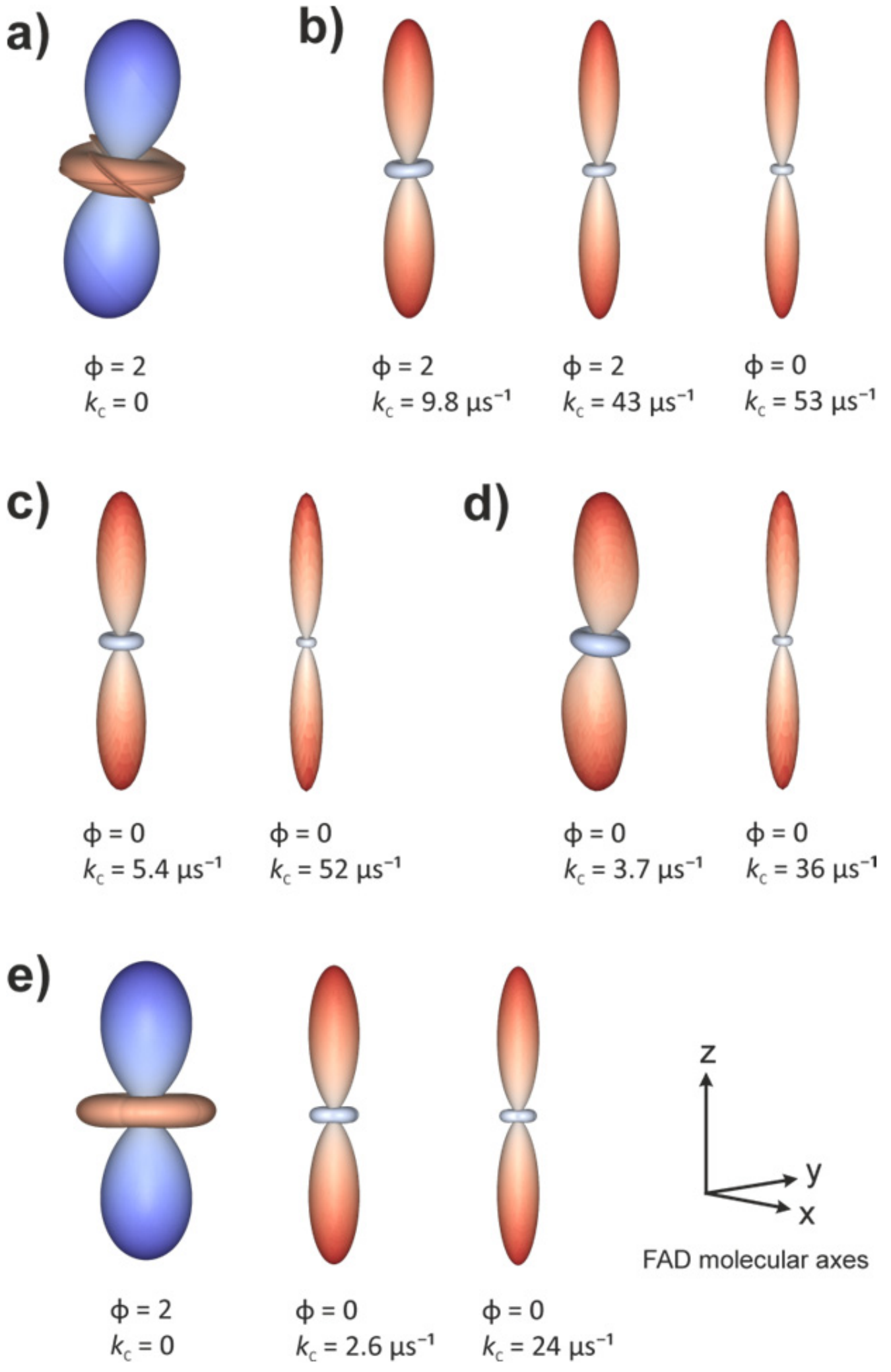

**Anisotropic yields of the signalling state, S. (**a) [$FAD^{\bullet-}$ $W^{\bullet+}$] radical pair with N5 and N10 in $FAD^{\bullet-}$ and N1 in $W^{\bullet+}$ and no scavenging reaction. (b), (c) and (d) show the results of the same calculation in the presence of scavenging by either (b) a radical with no hyperfine interactions, or (c) the ascorbyl radical model, or (d) another $FAD^{\bullet-}$ radical. (e) [$FAD^{\bullet-}$ $Z^{\bullet}$] radical pair with 7 nuclear spins in $FAD^{\bullet-}$ and none in $Z^{\bullet}$. Details of the model are discussed in the main text. The values of the parameters $k_C$ and $\phi$ were chosen so as to show the anisotropy in the absence of the scavenging reaction or the anisotropy corresponding to the maximum $\Delta_S$ or the maximum $\Gamma_S$ as a function of $k_C$ (see Figures 4 and 5). In these plots the distance in any direction from the centre of each pattern to the surface is proportional to $\left|Y_S(\Omega) - \langle Y_S \rangle\right|$ when the magnetic field has that direction. Red/blue regions correspond to reaction yields larger/smaller than the average.

## Figure S6

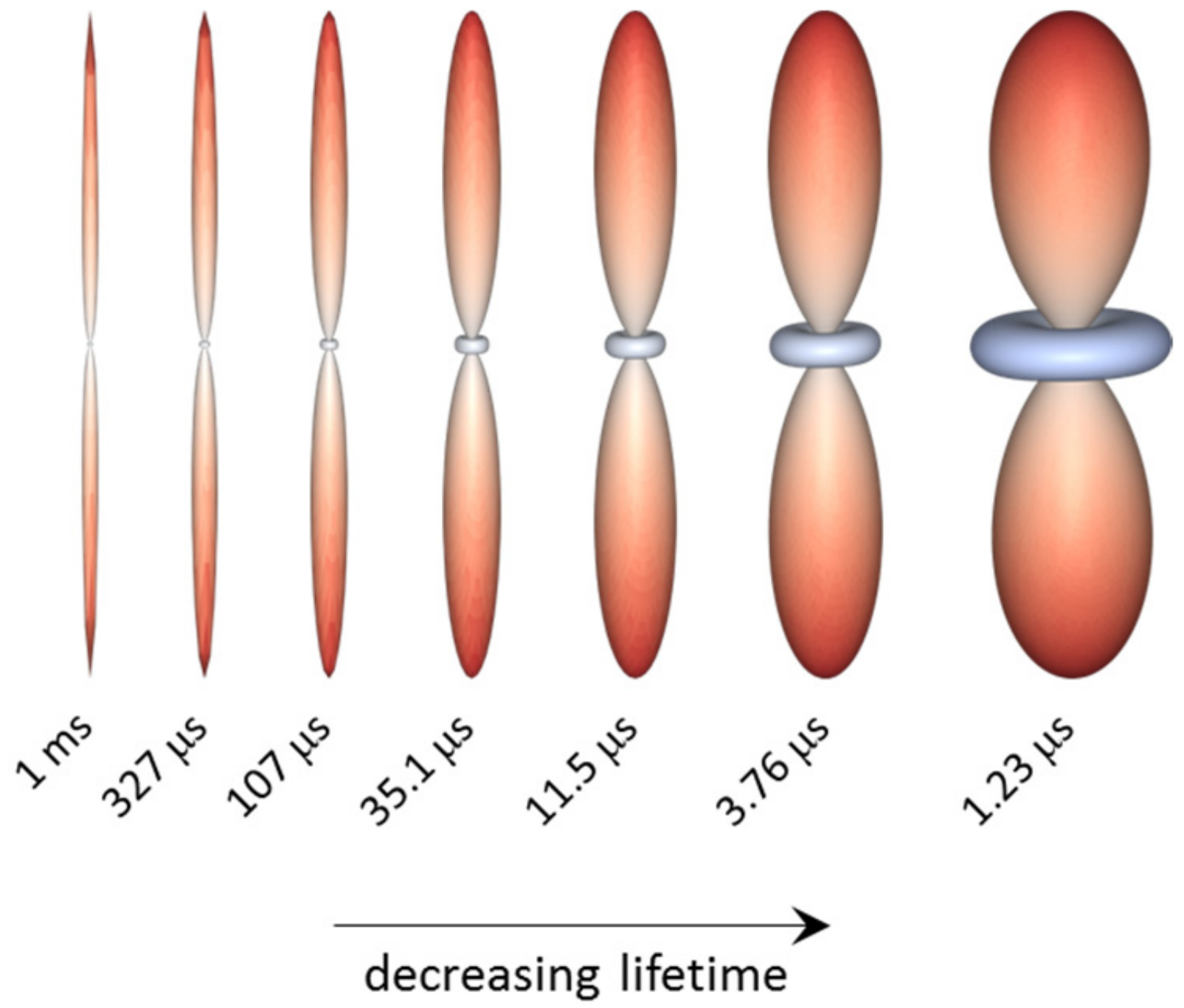

**Anisotropic yields of the signalling state, S.** [$FAD^{\bullet-}$ $WH^{\bullet+}$] radical pair with N5 and N10 in $FAD^{\bullet-}$ and N1 in $WH^{\bullet+}$. The scavenger, which reacted with $WH^{\bullet+}$, is a radical with no hyperfine interactions. $k_f = k_b$ with $k_f^{-1}$ = 1000, 327, 107, 35.1, 11.5, 3.76, and 1.23 µs as shown. $k_C$ = 65.1, 74.8, 79.2, 71.0, 49.6, 35.8, and 35.0 $\mu s^{-1}$, respectively, corresponding to the maximum anisotropy ($\Gamma_S$) for each value of $k_f$. The anisotropic yields were rescaled to reveal most clearly the increase in spikiness as the lifetime of the radical pair was prolonged. If drawn to scale, the pattern for $k_f^{-1}$ = 1 ms would be 64 times taller than that for $k_f^{-1}$ = 1.23 µs. See Figure S5 for a description of this type of plot.

## S1. Derivation of Eq. (3)

Consider a doublet ($S_A$ = ½) interacting with a particle with spin $S_C$ = $J$. According to the Clebsch-Gordan series, the tensor product states associated with $\hat{\mathbf{S}}_A$ and $\hat{\mathbf{S}}_C$ can be combined to give eigenstates of the total angular momentum $\hat{\mathbf{S}}_{AC} = \hat{\mathbf{S}}_A + \hat{\mathbf{S}}_C$ with quantum numbers $J \pm \frac{1}{2}$. The projection operator for the $J - \frac{1}{2}$ case is proportional to

$$\hat{P}_{AC}^{-} \propto \hat{S}_{AC}^{2} - (J + \tfrac{1}{2})(J + \tfrac{3}{2})\hat{1}, \tag{S.1}$$

because any state with the complementary total angular momentum quantum number $J + \frac{1}{2}$ will be an eigenstate of $\hat{S}_{AC}^{2}$ with eigenvalue $(J + \frac{1}{2})(J + \frac{3}{2})$ and thus annihilated by the term proportional to $\hat{1}$. Eq. (S.1) can be simplified by expanding $\hat{S}_{AC}^{2} = \left(\hat{\mathbf{S}}_A + \hat{\mathbf{S}}_C\right)^2 = \hat{S}_A^2 + \hat{S}_C^2 + 2\hat{\mathbf{S}}_A \cdot \hat{\mathbf{S}}_C$ and replacement of $\hat{S}_A^2$ and $\hat{S}_C^2$ by their respective eigenvalues multiplied by $\hat{1}$:

$$\hat{P}_{AC}^{-} = \frac{1}{N^{-}}\left(2\hat{\mathbf{S}}_A \cdot \hat{\mathbf{S}}_C - J\hat{1}\right). \tag{S.2}$$

Here, $N^{-}$ is a normalization constant to ensure that $\left(\hat{P}_{AC}^{-}\right)^2 = \hat{P}_{AC}^{-}$. An analogous argument suggests that

$$\hat{P}_{AC}^{+} = \frac{1}{N^{+}}\left(\hat{S}_{AC}^{2} - (J - \tfrac{1}{2})(J + \tfrac{1}{2})\hat{1}\right) = \frac{1}{N^{+}}\left(2\hat{\mathbf{S}}_A \cdot \hat{\mathbf{S}}_C + (J+1)\hat{1}\right), \tag{S.3}$$

The normalization constant can be established by requiring that $\hat{P}_{AC}^{+} + \hat{P}_{AC}^{-} = \hat{1}$, as the two total angular momentum states are mutually exclusive and complete. This yields

$$N^{+} = -N^{-} = 2J + 1 \tag{S.4}$$

which when combined with eqs (S.2) and (S.3) gives eq. (3).

## S2. Derivation of Eqs (6) and (8)

For the sake of clarity, we focus on the case $k_{\mathrm{C}}^{+}=0$. The general results given in eqs (6) and (8) may be derived in a similar fashion or obtained from the $k_{\mathrm{C}}^{+}=0$ result by substituting $k_{\mathrm{C}}^{-}\to k_{\mathrm{C}}^{-}-k_{\mathrm{C}}^{+}$ and multiplying the resulting expression by $\exp\left(-k_{\mathrm{C}}^{+}t\right)$.

Integrating eq. (4) for $k_{\mathrm{C}}^{+}=0$, we obtain:

$$\begin{aligned}\hat{\rho}(t) &= \frac{1}{\mathrm{Tr}\left[\hat{P}_{\mathrm{AB}}^{\mathrm{S}}\right]}\exp\left(-\tfrac{1}{2}k_{\mathrm{C}}^{-}\hat{P}_{\mathrm{AC}}^{-}\right)\hat{P}_{\mathrm{AB}}^{\mathrm{S}}\exp\left(-\tfrac{1}{2}k_{\mathrm{C}}^{-}\hat{P}_{\mathrm{AC}}^{-}\right)\\ &=\frac{1}{\mathrm{Tr}\left[\hat{P}_{\mathrm{AB}}^{\mathrm{S}}\right]}\exp\left(-\tfrac{1}{2J+1}k_{\mathrm{C}}^{-}\left(\tfrac{J}{2}-\hat{\mathbf{S}}_{\mathrm{A}}\cdot\hat{\mathbf{S}}_{C}\right)\right)\left(\tfrac{1}{4}-\hat{\mathbf{S}}_{\mathrm{A}}\cdot\hat{\mathbf{S}}_{\mathrm{B}}\right)\exp\left(-\tfrac{1}{2J+1}k_{\mathrm{C}}^{-}\left(\tfrac{J}{2}-\hat{\mathbf{S}}_{\mathrm{A}}\cdot\hat{\mathbf{S}}_{C}\right)\right).\end{aligned} \tag{S.5}$$

The exponential terms are diagonal in a coupled representation of $\hat{\mathbf{S}}_{\mathrm{A}}$ and $\hat{\mathbf{S}}_{\mathrm{C}}$, while the term derived from $\hat{\rho}(0)\propto\hat{P}_{\mathrm{AB}}^{\mathrm{S}}$ is diagonal in the coupled representation of $\hat{\mathbf{S}}_{\mathrm{A}}$ and $\hat{\mathbf{S}}_{\mathrm{B}}$. We evaluate $\mathrm{Tr}\left[\hat{\rho}(t)\right]$ in the basis of the *coupled* eigenstates of the total angular momentum of $\hat{\mathbf{S}}=\hat{\mathbf{S}}_{\mathrm{A}}+\hat{\mathbf{S}}_{\mathrm{B}}+\hat{\mathbf{S}}_{\mathrm{C}}$, $\hat{\mathbf{S}}_{\mathrm{AC}}=\hat{\mathbf{S}}_{\mathrm{A}}+\hat{\mathbf{S}}_{\mathrm{C}}$ and $\hat{\mathbf{S}}_{\mathrm{B}}$, i.e. the set of states $\left|((S_{\mathrm{A}},S_{\mathrm{C}})S_{\mathrm{AC}},S_{\mathrm{B}})S,M\right\rangle\equiv\left|(S_{\mathrm{AC}},S_{\mathrm{B}})S,M\right\rangle$. This yields

$$\hat{\rho}(t) = \frac{1}{\mathrm{Tr}\left[\hat{P}_{\mathrm{AB}}^{\mathrm{S}}\right]}\sum_{S_{AC}=J-\frac{1}{2}}^{J+\frac{1}{2}}\sum_{S=\left|S_{AC}-\frac{1}{2}\right|}^{S_{AC}+\frac{1}{2}}\sum_{M=-S}^{S}A^{2}(S_{AC})\left\langle(S_{AC},S_{B})S,M\right|\hat{P}_{\mathrm{AB}}^{\mathrm{S}}\left|(S_{AC},S_{B})S,M\right\rangle, \tag{S.6}$$

where

$$\begin{aligned}A(S_{AC})&=\left\langle(S_{AC},S_{B})S,M\right|\exp\left(-\tfrac{1}{2J+1}k_{\mathrm{C}}^{-}\left(\tfrac{J}{2}-\hat{\mathbf{S}}_{\mathrm{A}}\cdot\hat{\mathbf{S}}_{C}\right)\right)\left|(S_{AC},S_{B})S,M\right\rangle\\ &=\exp\left(-\tfrac{1}{4J+2}k_{\mathrm{C}}^{-}\left(J+S_{A}(S_{A}+1)+S_{C}(S_{C}+1)-S_{AC}(S_{AC}+1)\right)\right)\end{aligned} \tag{S.7}$$

which is independent of *S* and *M*. In order to evaluate the matrix elements of $\hat{P}_{\mathrm{AB}}^{\mathrm{S}}$, we recouple the angular momenta to yield an eigenbasis of $\hat{\mathbf{S}}_{AB}=\hat{\mathbf{S}}_{A}+\hat{\mathbf{S}}_{B}$, $\hat{\mathbf{S}}_{C}$, and $\hat{\mathbf{S}}$:

$$\left|((S_{A},S_{C})S_{AC},S_{B})S,M\right\rangle=\sum_{S_{AB}}\left|(S_{C},(S_{A},S_{B})S_{AB})S,M\right\rangle\left\langle(S_{C},(S_{A},S_{B})S_{AB})S\right|\left((S_{A},S_{C})S_{AC},S_{B})S\right\rangle, \tag{S.8}$$

with the recoupling coefficient given in terms of Wigner 6-j symbol by:

$$\left\langle(S_{C},(S_{A},S_{B})S_{AB})S\right|\left((S_{A},S_{C})S_{AC},S_{B})S\right\rangle=(-1)^{S_{A}+S_{B}+S_{C}+S}\sqrt{(2S_{AB}+1)(2S_{AC}+1)}\begin{Bmatrix}S_{C} & S_{A} & S_{AC}\\ S_{B} & S & S_{AB}\end{Bmatrix}. \tag{S.9}$$

Combining eq. (S.6) and (S.8), we may thus write

$$\hat{\rho}(t) = \frac{1}{\mathrm{Tr}\left[\hat{P}^{\mathrm{S}}_{\mathrm{AB}}\right]} \sum_{S_{AC}=J-\frac{1}{2}}^{J+\frac{1}{2}} \sum_{S=\left|S_{AC}-\frac{1}{2}\right|}^{S_{AC}+\frac{1}{2}} \sum_{S_{AB}=0}^{1} \sum_{M=-S}^{S} A^2(S_{AC})B(S_{AB}) \left|\left\langle (S_C, S_{AB})S \middle| (S_{AC}, S_B)S \right\rangle\right|^2$$
$$= \frac{1}{\mathrm{Tr}\left[\hat{P}^{\mathrm{S}}_{\mathrm{AB}}\right]} \sum_{S_{AC}=J-\frac{1}{2}}^{J+\frac{1}{2}} \sum_{S=\left|S_{AC}-\frac{1}{2}\right|}^{S_{AC}+\frac{1}{2}} \sum_{S_{AB}=0}^{1} (2S+1) A^2(S_{AC})B(S_{AB}) \left|\left\langle (S_C, S_{AB})S \middle| (S_{AC}, S_B)S \right\rangle\right|^2, \tag{S.10}$$

where

$$B(S_{AB}) = \left\langle (S_C, S_{AB})S, M \middle| \hat{P}^{\mathrm{S}}_{\mathrm{AB}} \middle| (S_C, S_{AB})S, M \right\rangle$$
$$= \frac{1}{2}\left(\tfrac{1}{2} + S_A(S_A+1) + S_B(S_B+1) - S_{AB}(S_{AB}+1)\right), \tag{S.11}$$

using the $M$-independence of the summands. It is clear that for the singlet initial configuration $B(S_{AB})$ vanishes except for the singlet basis, i.e. $S_{\mathrm{AB}} = 0$, for which $B(0) = 1$. This condition also implies that non-zero contributions can only result from $S = S_{\mathrm{C}} = J$. As a consequence,

$$\hat{\rho}(t) = \frac{(2J+1)}{\mathrm{Tr}\left[\hat{P}^{\mathrm{S}}_{\mathrm{AB}}\right]} \sum_{S_{AC}=J-\frac{1}{2}}^{J+\frac{1}{2}} A^2(S_{AC}) \left|\left\langle (S_C = J, S_{AB} = 0)S = J \middle| (S_{AC}, S_B = \tfrac{1}{2})S = J \right\rangle\right|^2$$
$$= \sum_{S_{AC}=J-\frac{1}{2}}^{J+\frac{1}{2}} A^2(S_{AC})(2S_{AC}+1) \left|\begin{Bmatrix} J & \frac{1}{2} & S_{AC} \\ \frac{1}{2} & J & 0 \end{Bmatrix}\right|^2. \tag{S.12}$$

Here, we have used $\mathrm{Tr}\left[\hat{P}^{\mathrm{S}}_{\mathrm{AB}}\right] = 2J+1$. The required 6-j symbol is

$$\left|\begin{Bmatrix} J & \frac{1}{2} & S_{AC} \\ \frac{1}{2} & J & 0 \end{Bmatrix}\right|^2 = \frac{1}{2(2J+1)}, \tag{S.13}$$

which allows us to evaluate the sum. Simple algebraic manipulation eventually yields eq. (6) for $k^{+}_{\mathrm{C}} = 0$.

An analogous approach can be used to derive the singlet probability in the subspace of spin A and B (eq. (8)). In particular,

$$\mathrm{Tr}\left[\hat{P}^{\mathrm{S}}_{\mathrm{AB}}\,\hat{\rho}(t)\right] = \frac{1}{\mathrm{Tr}\left[\hat{P}^{\mathrm{S}}_{\mathrm{AB}}\right]} \mathrm{Tr}\left[\hat{P}^{\mathrm{S}}_{\mathrm{AB}} \exp\left(-\tfrac{1}{2}k^{-}_{\mathrm{C}}\hat{P}^{-}_{\mathrm{AC}}\right) \hat{P}^{\mathrm{S}}_{\mathrm{AB}} \exp\left(-\tfrac{1}{2}k^{-}_{\mathrm{C}}\hat{P}^{-}_{\mathrm{AC}}\right)\right]$$
$$= \sum_{S_{AC}=J-\frac{1}{2}}^{J+\frac{1}{2}} \sum_{S'_{AC}=J-\frac{1}{2}}^{J+\frac{1}{2}} A(S_{AC})A(S'_{AC}) \left|\left\langle (S_C = J, S_{AB} = 0)S = J \middle| (S_{AC}, S_B = \tfrac{1}{2})S = J \right\rangle\right|^2 \cdot$$
$$\left|\left\langle (S_C = J, S_{AB} = 0)S = J \middle| (S'_{AC}, S_B = \tfrac{1}{2})S = J \right\rangle\right|^2, \tag{S.14}$$

from which eq. (8) follows.